\documentclass[review]{elsarticle}
\journal{International Journal of Multiphase Flow}

\usepackage{color}
\usepackage{multirow}
\usepackage{dcolumn}
\usepackage{multicol}
\usepackage{graphicx}
\usepackage{caption}
\usepackage{amssymb}
\usepackage{amsmath}
\usepackage{epstopdf}
\usepackage{mathtools}
\usepackage{float}
\usepackage{bm}

\usepackage{subcaption}
\usepackage{graphicx}
\usepackage{soul}

\biboptions{authoryear}

\biboptions{sort&compress}

\newcommand{\ks}{\textcolor{black}} 

\begin{document}

\begin{frontmatter}
\title{Instabilities in Drying Colloidal Films: Role of Surface Charge and Substrate Wettability}
\author{A. Madhav Sai Kumar$^a$, A. Hari Govindha$^b$, Ranajit Mondal$^a$ and \\ Kirti Chandra Sahu$^{c}$\footnote{ksahu@che.iith.ac.in} }
\address{$^a$Soft Matter Group, Department of Chemical Engineering, Indian Institute of Technology Hyderabad, Kandi 502 284, Telangana, India \\
$^b$Department of Mechanical and Aerospace Engineering, Indian Institute of Technology Hyderabad, Kandi - 502 284, Sangareddy, Telangana, India  \\ 
$^c$Department of Chemical Engineering, Indian Institute of Technology Hyderabad, Kandi - 502 284, Sangareddy, Telangana, India}

\begin{abstract}
The drying of colloidal suspensions leads to complex deposition patterns, accompanied by instabilities such as cracking and delamination. In this study, we experimentally investigate the coupled influence of particle surface charge and substrate wettability on the evaporation dynamics, final deposition morphology, and crack patterns of sessile droplets containing silica nanoparticles. We examine the dynamics of two types of colloids, namely the negatively charged colloidal silica nanoparticles (Ludox TM50) and the positively charged silica nanoparticle (Ludox CL30), at concentrations ranging from 0.1 wt.\% to 5.0 wt.\%, deposited on glass, polystyrene, and polytetrafluoroethylene (PTFE) substrates with distinct wettability. Side and top-view imaging techniques are employed to capture the evaporation process and analyze the resulting cracks. Our results reveal that the nature of the particle charge and substrate wettability significantly affect the evaporation mode, with transitions observed between constant contact radius (CCR), constant contact angle (CCA), and mixed modes. TM50-laden droplets consistently exhibit radial cracks, whereas CL30 droplets display more randomly oriented and irregular cracks. At higher particle concentrations, TM50 suspensions form thicker deposits that undergo delamination, particularly on highly wettable substrates like glass. Quantitative analysis reveals that crack spacing and length follow power-law relationships with particle concentration. Additionally, the delamination behavior is strongly influenced by both the particle concentration and the type of substrate. We propose a mechanistic framework to explain the role of particle–substrate interactions in governing the observed cracking and delamination behaviors.
\end{abstract}
\end{frontmatter}


\section{Introduction} \label{sec:intro}

The evaporation of complex fluids, such as colloidal dispersions, polymer solutions, and biological fluids, involves the irreversible removal of solvent, leading to the consolidation of solutes into dense deposits. This drying process plays a vital role in diverse applications, including paints and coatings \citep{brinckmann2011experimental}, cosmetics \citep{yu2021droplet}, microelectronics \citep{davoust2013evaporation, prasad2014monitoring}, inkjet printing \citep{de2004inkjet, lim2012deposit, lim2009experimental}, and lithography \citep{kim2005importance, he2019evaporation, chang2006evaporation, kim2016controlled, yanagisawa2014investigation,weldon2017uniformly}. 

The evaporation of particle-laden dispersions has been extensively studied in the sessile configuration, where it often results in a ring-like deposit of particles at the periphery, a phenomenon known as the “coffee-ring” effect \citep{deegan1997capillary}. This pattern arises from outward capillary flows induced by a non-uniform evaporative flux across the droplet surface \citep{deegan1997capillary, karapetsas2016evaporation, katre2021evaporation, katre2022experimental}. Over the past few decades, significant progress has been made in understanding and controlling the evaporative assembly of colloidal particles into well-defined patterns, supported by theoretical predictions of internal flow and solute transport within evaporating droplets that demonstrated good agreement with experimental observations \citep{deegan2000contact}. Both extrinsic factors, such as substrate properties \citep{mondal2018patterns}, drying configuration \citep{mondal2018patterns, mondal2019influence, mondal2020patterning}, temperature \citep{li2015coffee}, and humidity \citep{zhang2022ultrafast}, and intrinsic parameters, including particle size \citep{weon2010capillary}, shape \citep{yunker2011suppression, dugyala2014control,hodges2010influence}, surface charge \citep{dugyala2014control, bhardwaj2010self}, particle density \citep{liou2024suppression}, dispersing medium \citep{katre2021evaporation,katre2020evaporation,hari2022counter,gurrala2019evaporation}, and the presence of macromolecules \citep{shao2020role, still2012surfactant, rey2022versatile} or small dissolved species \citep{ma2012effect, xu2017effect, shimobayashi2018suppression}, have been shown to play crucial roles in dictating particle transport and deposition. These factors collectively influence the formation of intriguing deposit patterns \citep{mondal2023physics}. A detailed review of the drying dynamics of colloidal droplets is available in \citet{thampi2023drying}.

The drying of colloidal dispersions that form particulate films on substrates often leads to mechanical defects such as buckling, wrinkling, cracking, warping, or delamination \citep{mondal2023physics,lohani2020coupled,lama2017tailoring}. While such defects are generally undesirable in applications like paints and coatings, controlled crack formation with designed periodicity holds promise for optical grafting and as lithographic templates for fabricating nano- and microfluidic channels. Fundamentally, these defects originate from the release of excess strain energy accumulated during the consolidation of particles as the solvent evaporates \citep{dugyala2016role}.

Numerous studies have investigated crack formation and evolution in drying colloidal films, emphasizing the influence of material properties \citep{mailer2014cracking,dugyala2016role} and external conditions \citep{lama2016magnetic,lama2017tailoring,lama2018cracks}. For instance, \citet{giorgiutti2014elapsed} experimentally explored the relationship between spatial ordering and cracking onset to the mechanical properties of the gel and drying kinetics using sessile droplets of aqueous colloidal silica on glass substrates. \citet{dugyala2016role} demonstrated that particle shape significantly influences crack morphology. Specifically, they observed that spherical particles tend to form radial cracks, while ellipsoids lead to circular crack patterns. This transition reflects differences in particle self-assembly and local ordering, with cracks propagating along paths of least resistance to release the accumulated strain energy. A hydrodynamic model proposed by \citet{lilin2022criteria} linked the pore pressure within the deposit to the initiation of cracks and air invasion during drying, and was validated through experiments involving the evaporation of Ludox AS30 particles on glass. The study identified two distinct air invasion modes, governed by the interplay of pore pressure and capillary forces, which modulate internal stress distributions. Substrate stiffness also plays a key role, with softer substrates tending to produce circular cracks, while stiffer ones favor linear or radial crack patterns \citep{lama2021synergy}. Furthermore, \citet{kumar2023effect} showed that particle polarity and surface charge impact crack spacing and propagation speed, with positively charged particles yielding wider and faster-growing cracks due to changes in adhesion and interparticle interactions. 

\ks{Surface charge plays a crucial role in governing film cracking behavior. By varying the pH of the suspension, \citet{singh2009cracking} experimentally examined the influence of zeta potential on the critical cracking thickness (CCT) of aqueous $\alpha$-alumina dispersions deposited on glass substrates. They observed that the CCT increases as the zeta potential approaches zero and subsequently decreases with further reduction in zeta potential. This non-monotonic behavior was attributed to changes in surface charge that alter the final particle packing fraction in the dried film and, consequently, the CCT. The importance of particle–substrate interactions was further highlighted by \citet{ghosh2015effect}, who investigated the cracking dynamics and morphology of colloidal films composed of polystyrene particles with a surface potential of 15~mV deposited on glass substrates with varying wettabilities and surface potentials of $-30$~mV, $-40$~mV, and $-89$~mV. Their results demonstrated that increasing substrate hydrophobicity weakens particle–substrate attractive forces, resulting in a higher threshold stress for crack initiation, greater resistance to rupture, larger crack openings, and a reduced crack propagation velocity. In a related study, \citet{yan2008particle} systematically modified the surface charges of both polystyrene particles and glass substrates. When particles and substrates carried the same charge, ordered deposition patterns emerged, accompanied by reduced adhesion and enhanced particle mobility. In contrast, oppositely charged particle–substrate systems produced disordered deposits characterized by strong adhesion and limited particle mobility. More recently, \citet{bridonneau2020self} examined particle self-assembly driven by particle–substrate interactions using silicon nanoparticles deposited on gold substrates with controlled surface potentials. They reported that although particle concentration strongly influences the resulting deposit morphology, the overall deposit shape remains largely insensitive to the specific particle–substrate interaction.}

The influence of environmental conditions on deposition patterns and crack morphology has also been explored. \citet{lama2017tailoring} demonstrated that silica nanoparticle droplets dried on heated glass substrates formed more ordered crack patterns compared to those dried at ambient temperature, resulting from modified solvent flow and capillary transport. \citet{chhasatia2010effect} observed that the higher relative humidity suppressed coffee-ring formation by slowing evaporation and reducing the contact angle, which led to broader droplet spreading. \citet{bourrianne2021crack} identified regimes of non-uniform deposit thickness associated with diverse crack morphology and observed that, in some cases, the deposit delaminated from the substrate due to avalanche-like crack propagation, depending on the initial particle concentration. Both the deposit geometry and crack density were found to be highly sensitive to the particle volume fraction. Furthermore, \citet{sanyal2015agglomeration} reported two distinct drying regimes in water droplets containing silica nanoparticles. The contact line remained pinned in the first regime, whereas in the second regime, both the contact line and the agglomeration front receded inward. This transition resulted in the formation of radial cracks that propagated inward from the edge of the droplet. 

Several studies have investigated crack formation in thin films, emphasizing the roles of material composition, film thickness, and drying conditions. \citet{giorgiutti2016painting} examined crack patterns in colloidal films on non-porous substrates and proposed a non-invasive method to assess the mechanical properties of painted surfaces, particularly for heritage conservation applications. \citet{bamboriya2025effective} studied films composed of hard–soft colloid mixtures, showing that cracking depends on particle stiffness, packing, and film thickness. They identified a critical thickness for crack-free films and linked the elastic modulus to the particle ratio, with deviations attributed to the sedimentation of denser particles. \citet{goehring2013evolving} classified crack geometries in natural and synthetic systems as rectilinear (right-angle intersections) or hexagonal ($120^\circ$ intersections), with transitions observed during repeated healing cycles. \citet{nam2012patterning} demonstrated controlled crack patterning in silicon nitride films on silicon substrates, identifying three morphologies, namely straight, oscillatory, and bifurcated, by varying processing conditions and film parameters. \citet{tirumkudulu2005cracking} explored the influence of particle size in latex films, finding that larger particles yielded three visually distinct domains, whereas smaller particles formed transparent films with different cracking behavior. \citet{groisman1994experimental} observed a linear relationship between crack spacing and film thickness in coffee–water mixtures, with cracks propagating vertically through the film depth.

Beyond crack propagation, buckling and delamination are also critical phenomena observed during the drying of colloidal deposits, having significant implications for various applications. \citet{lohani2020coupled} showed that delamination is more likely on silicon wafers at higher particle concentrations and smaller particle sizes, especially on rough substrates. \citet{osman2020controlling} demonstrated that the addition of polyethylene oxide suppresses delamination by modifying suspension properties and slowing drying kinetics. In a related study,  \citet{pauchard2004invagination} reported that drying latex suspensions on superhydrophobic substrates led to a transition from rigid, shell-like deposits to toroidal structures as particle concentration decreased; intermediate concentrations produced partially invaginated shells. \citet{lama2023salinity} found that increasing ionic strength enhances particle adhesion and film cohesion while reducing bending, consistent with a theoretical model that attributes bending to uneven shrinkage. More recently, \citet{liu2024formation} identified a critical deposit thickness during the drying of silica-laden droplets, which governs the transition from purely radial cracks to a combination of radial and circular cracks, influenced by substrate stiffness.

As highlighted in the preceding discussion, the evaporation of particle-laden droplets has attracted considerable interest, with numerous studies investigating how parameters such as substrate wettability, ambient conditions, interparticle interactions, and electrostatic effects influence particle transport and deposition. These factors significantly affect the internal flow dynamics within the droplet and ultimately govern the final deposition morphology and crack formation mechanisms. Unlike earlier studies that examined these effects in isolation, the present work focuses on the coupled influence of particle surface charge and substrate wettability. By systematically varying these parameters under identical experimental conditions, we investigate their combined influence on evaporation dynamics, final deposit morphology, and the extent of crack formation across different substrate types. This integrated approach enables more comprehensive and meaningful comparisons, offering new insights into the interdependent mechanisms driving pattern formation in drying colloidal droplets. In the present study, sessile evaporation experiments are conducted using two types of colloidal silica particles (Ludox TM50 and CL30 obtained from Sigma–Aldrich) on substrates with varying wettability, including glass, polystyrene, and Polytetrafluoroethylene (PTFE). Droplets of 2 $\mu$l volume, with particle concentrations ranging from 0.1 wt.\% to 5.0 wt.\%, are deposited and allowed to dry under controlled conditions. The drying process is captured using a Dino-Lite digital microscope and CMOS cameras from both top and side views. For Ludox TM50 droplets dried on glass substrates, classic coffee-ring patterns form at concentrations below 2.0 wt.\%, with cracks confined to the annular ring. At higher concentrations (3.0 wt.\% and 5.0 wt.\%), cracks propagate inward toward the center, accompanied by film wrinkling or delamination. In contrast, Ludox CL30 droplets also exhibit coffee-ring patterns, but random cracks emerge at lower concentrations, and at higher concentrations, these cracks extend toward the droplet center. Furthermore, a mechanistic framework is proposed to elucidate the roles of capillary forces, interfacial interactions, and particle–substrate forces (both repulsive and attractive) in governing the cracking and delamination observed during drying.

The remainder of the paper is organized as follows. Section \S\ref{sec:expt} outlines the experimental setup and procedure. The results are presented in Section \S\ref{sec:results}, where we analyze the evaporation kinetics, including the temporal evolution of wetting diameter, height, contact angle, and droplet volume during the drying process. Additionally, we examine the cracks formed in the final deposition patterns and evaluate the extent of film delamination associated with higher concentrations. The conclusions are presented in Section \S\ref{sec:conc}.

\section{Methodology} \label{sec:expt}

\subsection{Materials} 

We consider two types of model colloidal silica suspensions obtained from Sigma-Aldrich. They are negatively charged Ludox TM50 (diameter $d = 25$ nm; zeta potential, $\zeta = -33$ mV) and positively charged Ludox CL30 ($d = 12$ nm, $\zeta = 44$ mV). The properties of these nanoparticles are listed in Supplementary Table T1. We prepare colloidal suspensions of these nanoparticles at various concentrations ($\phi$) ranging from 0.1 wt.\% to 5.0 wt.\% by diluting the stock solutions of Ludox TM50 and Ludox CL30 with deionized water. The drying of nanoparticle-laden solutions at various concentrations is examined on three different substrates with varying wettabilities. The substrates used were hydrophilic glass (obtained from Blue Star Ltd.) with a water contact angle of $\theta = 22 \pm 2^\circ$, polystyrene with $\theta = 77 \pm 5^\circ$, and hydrophobic PTFE substrate (obtained from Saint-Gobain) with $\theta = 102 \pm 2^\circ$. The cleaning procedure of the substrates involved sequential washing with tap water, deionized water, and acetone, followed by air drying. The prepared nanoparticle suspensions were sonicated and then mixed using a vortex shaker to ensure homogeneous dispersion prior to the drying experiments. In each experiment, a sessile droplet of $2~\mu$l was carefully deposited on the substrates using a micropipette (range: $0-2.5~\mu$l). For each condition, a minimum of three repetitions was performed to ensure repeatability. All experiments were carried out at an ambient temperature of $24.0 \pm 1.0^\circ$C and a relative humidity of $58\pm10~\%$.

\subsection{Experimental procedure}  

\begin{figure}
\centering
\includegraphics[width=0.9\textwidth]{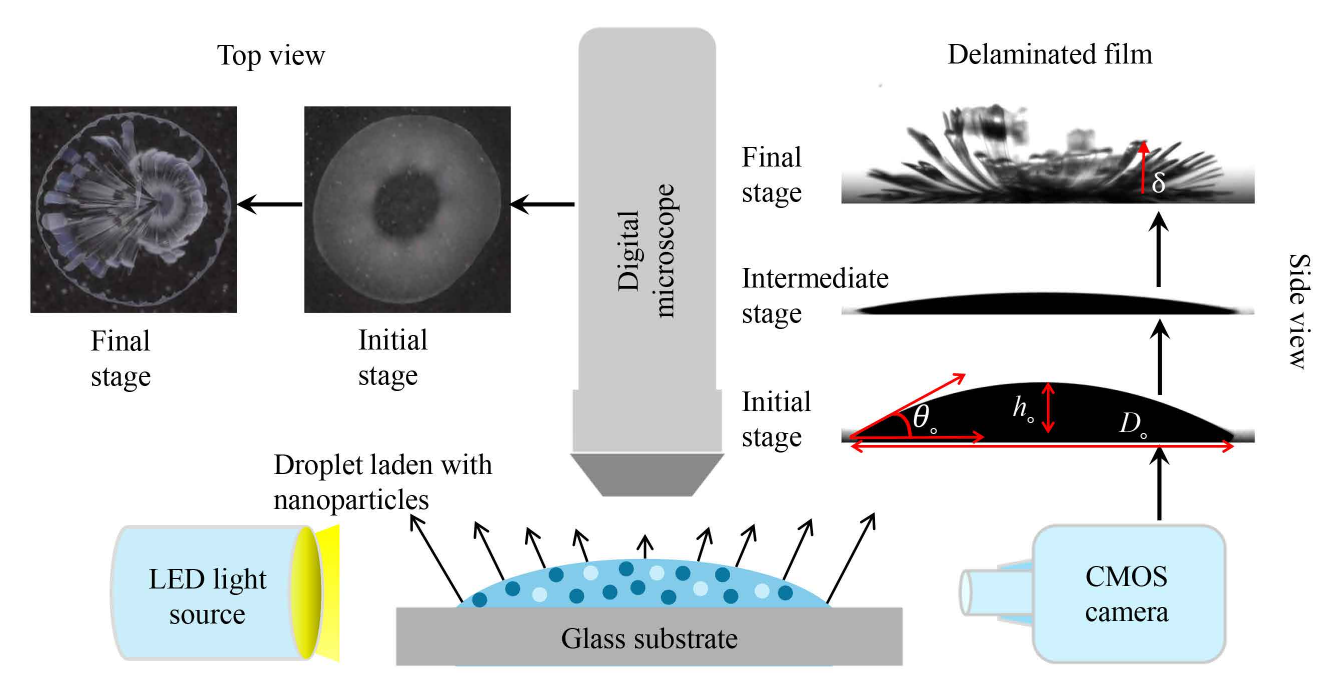}
\caption{Schematic of the experimental setup (customized goniometer for side-view imaging and digital microscope for top-view imaging) used to study the evaporation kinetics of nanoparticle-laden droplets. The top views recorded using digital microscopy and the side views of the droplet captured with the CMOS camera at different stages of evaporation are shown in the left and right insets of the figure, respectively.} 
\label{fig:fig1}
\end{figure}

Shadowgraphy and digital microscopy techniques were employed to investigate the evaporation kinetics of sessile water droplets laden with silica nanoparticles. A schematic of the experimental setup is provided in Fig.~\ref{fig:fig1}. For side-view imaging via shadowgraphy, a complementary metal-oxide-semiconductor (CMOS) camera (Make: Do3Think, Model: DS-CBY501E-H; resolution: $2592 \times 1944$ pixels) was used in conjunction with a light-emitting diode (LED) source. A diffuser sheet was placed in front of the LED to ensure uniform backlighting across the field of view of the camera. To capture the top view of the drying droplet, a digital microscope (Dino-Lite, AnMo Electronics Corp., Taiwan) with $50\times$ magnification was used, recording at 30 frames per second (fps). The microscope includes an adjustable LED illumination system to maintain consistent visibility under different ambient lighting conditions. Relative humidity was continuously monitored using a hygrometer (Make: HTC, Model: 288-ATH). The entire setup was enclosed within a goniometer box to minimize external disturbances.

\subsection{Post-processing}

The images of the droplet undergoing evaporation and drying, recorded using digital microscopy and a CMOS camera, were post-processed to extract the instantaneous height $(h)$, wetted diameter $(D)$, and contact angle $(\theta)$ during evaporation, as well as to analyze crack patterns and film wrinkling or delamination. An in-house solver was developed within the \textsc{Matlab}$^{\circledR}$ framework to process side-view images captured by the CMOS camera and extract droplet profiles. The processing workflow involved image sharpening and edge enhancement using an unsharp masking technique, followed by median filtering to eliminate random noise. The images were then binarized using a custom filter to isolate the droplet boundary from the background. Reflections were masked, and any discontinuities in the droplet outline were filled. The resulting binary contour was traced to obtain time-resolved measurements of droplet diameter, height, and contact angle throughout the evaporation process. Further details of the image processing methodology are available in \citet{katre2021evaporation}.

Top-view images were analyzed using Dino Capture 2.0 software (version 1.5.48.B) to measure the droplet diameter, crack spacing or width ($\lambda$), and crack length ($l$) in the final deposition patterns \citep{mondal2018patterns}. Accurate measurements were ensured by calibrating the built-in measurement tools with the appropriate magnification settings of the digital microscope. Additionally, ImageJ software with in \textsc{Matlab}$^{\circledR}$ framework was employed to quantify the degree of delamination ($\delta$) of the particulate film over time, particularly for higher particle concentrations ($\phi = 3$ wt\% and 5 wt\%).

\section{Results and Discussion}\label{sec:results}

\subsection{Deposition and crack Patterns}

As discussed in the introduction, the type of nanoparticles plays a crucial role in droplet evaporation and the resulting deposition patterns due to differences in particle size, shape, surface chemistry, and their interactions with both the solvent and the substrate. In this study, we examine how particle concentration, surface charge, and substrate wettability influence the deposition morphology and crack formation. In the following, we systematically compare the deposition patterns formed by Ludox TM50 (negatively charged) and Ludox CL30 (positively charged) on substrates with different wettability, while varying the particle concentrations from 0.1 wt.\% to 5.0 wt.\%.

\begin{figure}
\centering
\includegraphics[width=0.8\textwidth]{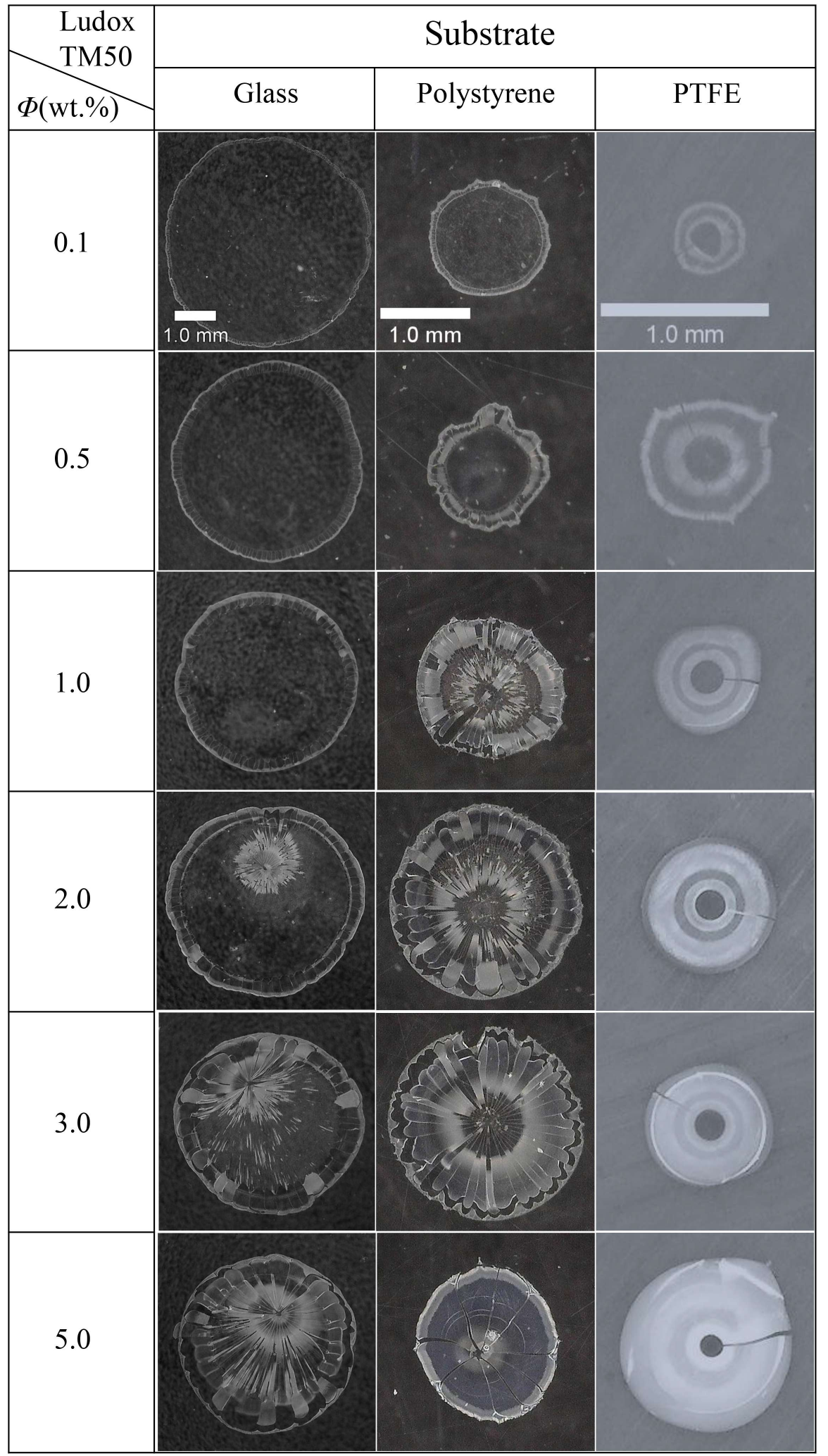}
\caption{Final deposition patterns of Ludox TM50 at different concentrations ($\phi$) on glass, polystyrene, and PTFE substrates.}
\label{fig:fig2}
\end{figure}

\begin{figure}
\centering
\includegraphics[width=0.8\textwidth]{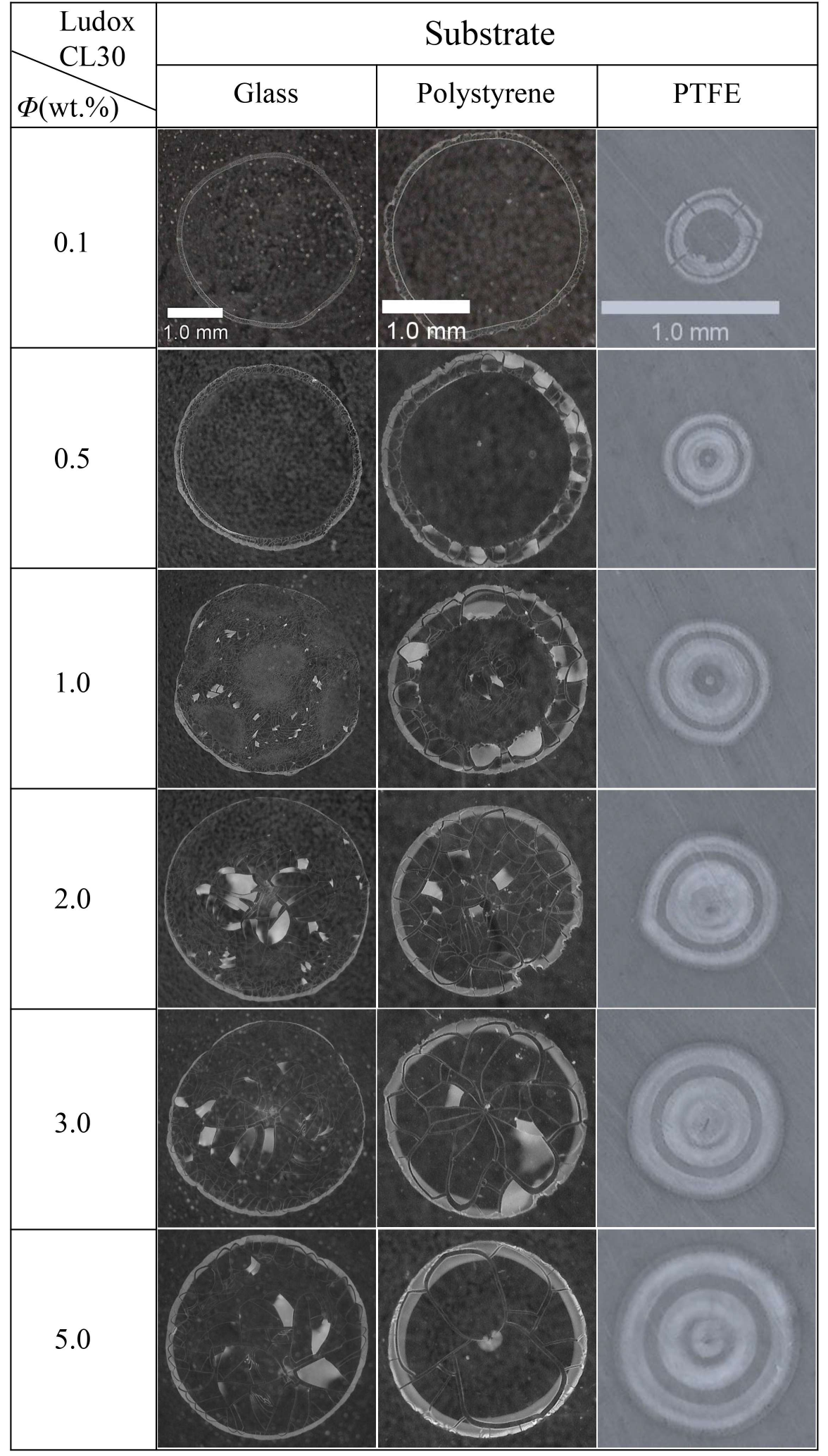}
\caption{Final deposition patterns of Ludox CL30 at different concentrations ($\phi$) on glass, polystyrene, and PTFE substrates.}
\label{fig:fig3}
\end{figure}

Figures~\ref{fig:fig2} and \ref{fig:fig3} depict the final dried deposition patterns resulting from the evaporation of $2~\mu$l sessile droplets containing Ludox TM50 and CL30 nanoparticles, respectively. The experiments are conducted on glass, polystyrene, and PTFE substrates, with particle concentrations varied as $\phi = 0.1$, 0.5, 1.0, 2.0, 3.0, and 5.0 wt.\%. The initial contact angle $(\theta_0)$ of droplets containing Ludox TM50 and CL30 nanoparticles on glass, polystyrene, and PTFE substrates is measured to be $22 \pm 2^\circ$, $77 \pm 5^\circ$, and $102 \pm 2^\circ$, respectively. We observe that the initial contact angle ($\theta_0$) remains nearly constant across different particle concentrations and types, indicating that $\theta_0$ is primarily determined by the substrate wettability rather than the particle concentration within the range considered in this study.

At the lowest concentration ($\phi = 0.1$ wt.\%), a thin coffee-ring-like deposit forms, characterized by a dense network of radial cracks along the periphery. As the concentration increases to $\phi = 0.5$ wt.\%, the ring becomes slightly wider and the number of radial cracks decreases. With further increase in concentration, both the ring width and crack extent grow, with cracks beginning to propagate inward toward the droplet center. Interestingly, for $\phi \geq 2$ wt.\%, the number of cracks reduces compared to the droplet with 0.5 wt.\% concentration, though the cracks extend deeper into the deposit. These observations are consistent with the findings of \citet{dugyala2016role}, who demonstrated that sessile droplets containing spherical particles typically develop radial crack patterns. At higher concentrations ($\phi \ge 3.0$ wt.\%), the cracks reach the central region of the deposit and often resulting in delamination of the colloidal film from the substrate. Such delamination behavior is also observed by \citet{liu2024formation,lilin2022criteria,sanyal2015agglomeration} during the drying of silica suspensions on glass substrates.

Next, we examine the drying behavior of Ludox CL30 (positively charged) droplets. The effect of particle concentration on the final deposition patterns formed on the glass substrate is shown in Fig.~\ref{fig:fig3} (left panel). At lower concentrations (0.1 wt.\% to 0.5 wt.\%), Ludox CL30 droplets form coffee-ring-like deposits, similar to those observed for Ludox TM50, but with random cracks distributed along the ring circumference. As the concentration increases from 1.0 wt.\% to 5.0 wt.\%, the cracks begin to extend inward, progressively covering a larger portion of the deposit. Similar deposition patterns were reported by \citet{kumar2023effect}. Unlike Ludox TM50, CL30 droplets exhibit two distinct types of cracks: radial cracks and transversely connecting cracks, oriented approximately perpendicular to each other. This difference may arise from electrostatic interactions between the positively charged CL30 particles and the negatively charged glass substrate \citep{somasundaran1995deposition}. The attractive interaction between the substrate and the particles leads to uneven or random particle settlement, increasing pore formation and promoting irregular crack propagation \citep{kumar2023effect}. Additionally, at higher concentrations, broader cracks are observed, as shown in Fig.~\ref{fig:fig3} (left panel).

To examine the effect of substrate wettability on the final deposit and crack patterns, Ludox TM50 and Ludox CL30 suspensions are dried on polystyrene and PTFE substrates at various particle concentrations. The resulting deposition patterns are shown in the middle and right panels of Fig.~\ref{fig:fig2} (Ludox TM50 droplet) and Fig. \ref{fig:fig3} (Ludox CL30 droplet). In Fig.~\ref{fig:fig2} (middle panel), corresponding to the polystyrene substrate, the deposition pattern at 0.1 wt.\% Ludox TM50 exhibits a classical coffee-ring pattern with radial cracks along the periphery. As the concentration increases to 0.5 wt.\%, the number of cracks decreases, while the ring becomes noticeably wider. For concentrations between 1.0 wt.\% and 5.0 wt.\%, the cracks extend inward toward the droplet center and eventually lead to delamination of the colloidal film. This behavior is similar to that observed on glass substrates at 3.0 wt.\% and 5.0 wt.\% Ludox TM50 concentrations. On the PTFE substrate, as shown in Fig.~\ref{fig:fig2} (right panel), the deposits at lower concentrations (0.1 wt.\% and 0.5 wt.\%) primarily consist of particles accumulating near the periphery and inner region, accompanied by the formation of two to three radial cracks. As the particle concentration increases from 1.0 wt.\% to 5.0 wt.\%, the resulting deposits become thicker and more uniform, typically featuring a single prominent crack that extends toward the center. At the highest concentrations, the deposits exhibit a doughnut-like morphology, characteristic of buckling-induced deformation.

The final deposition patterns of Ludox CL30 at various particle concentrations ranging from 0.1 wt.$\%  - 5.0 $ wt.\%  dried on the polystyrene substrate can be seen in  Fig. \ref{fig:fig3} (middle panel). At lower concentrations (i.e., 0.1 wt.\% and 0.5 wt.\%), the deposit consists of a ring with random cracks along its circumference, accompanied by an increase in ring width. As the concentration is increased further to 1.0 wt.$\%$, the cracks are observed to start forming in the central part. With further increase in concentration of particles to 2.0, 3.0 and 5.0 wt.$\%$, the cracks observe to cover the entire deposit alongside with the reduction of the number of cracks with larger crack spacing. In contrast, when the droplet containing Ludox CL30 dried on PTFE substrate, two to three radial cracks are formed at a very low particle concentration i.e., 0.1 wt.$\%$ and no crack in the cases of higher particle concentrations. In the case of lower concentrations of Ludox CL30 from 0.1 wt.$\%  - 1.0 $ wt.\%, there is a formation of two concentric rings, one at the periphery forming a ring and another in the interior region with no particles in the central region. As the concentration of ludox increased to 2.0 wt.\% and higher, the particles settled more in the central region and remained at the periphery, forming a thicker central ring and a thinner ring at the periphery, as shown in the Fig. \ref{fig:fig3} (right panel). To gain insight into distinct deposition patterns resulting from various combinations of operating parameters, we analyze the drying kinetics in the following section.

\subsection{Evaporation dynamics}

\begin{figure}
\centering
\hspace{0.6cm}{\large (a)} \hspace{4.5cm}{\large (b)} \\
 \includegraphics[width=0.95\textwidth]{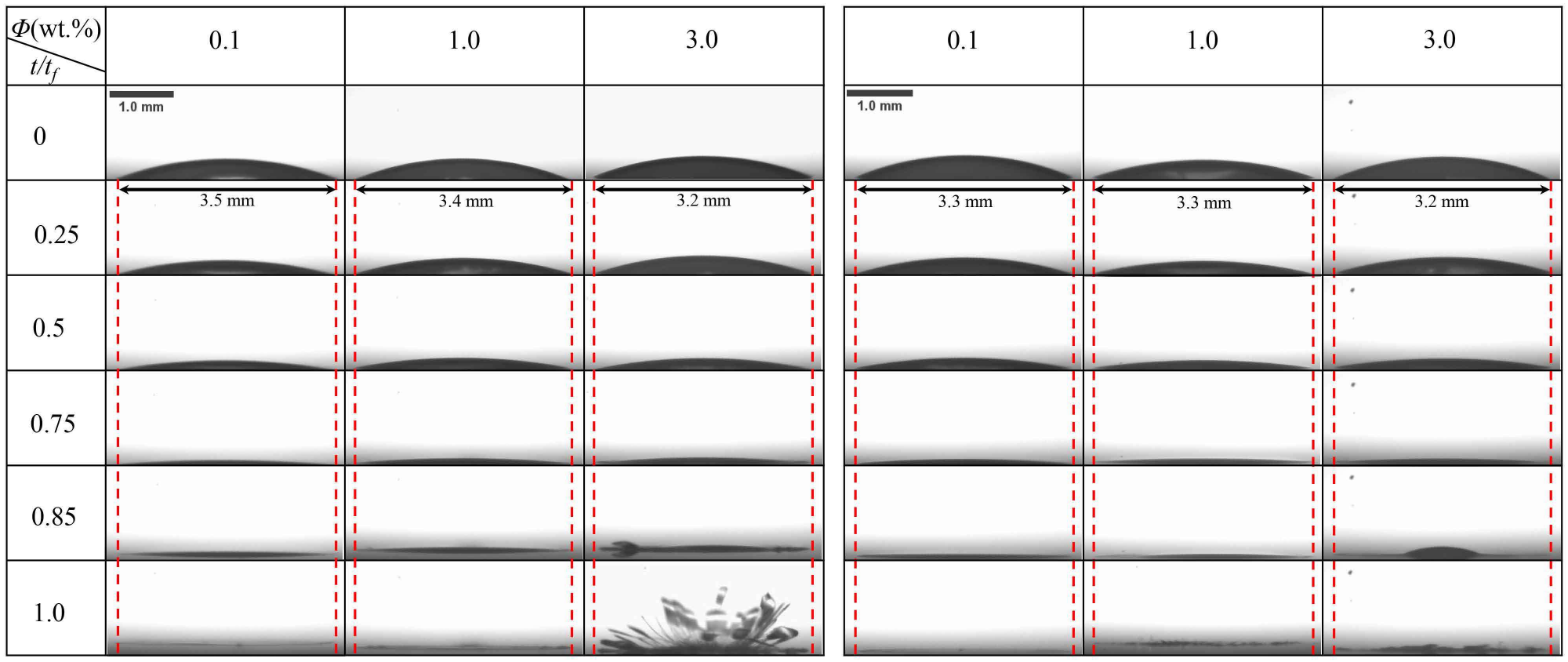}
 \caption{Temporal evolution (side view) of (a) Ludox TM50 and (b) Ludox CL30 sessile droplets at different concentrations $\phi$ (0.1 wt.\%, 1.0 wt.\%, and 3.0 wt.\%) on a glass substrate.} 
\label{fig:fig4}
\end{figure} 

\begin{figure}
\centering
\hspace{0.6cm}{\large (a)} \hspace{4.5cm}{\large (b)} \\
\includegraphics[width=0.95\textwidth]{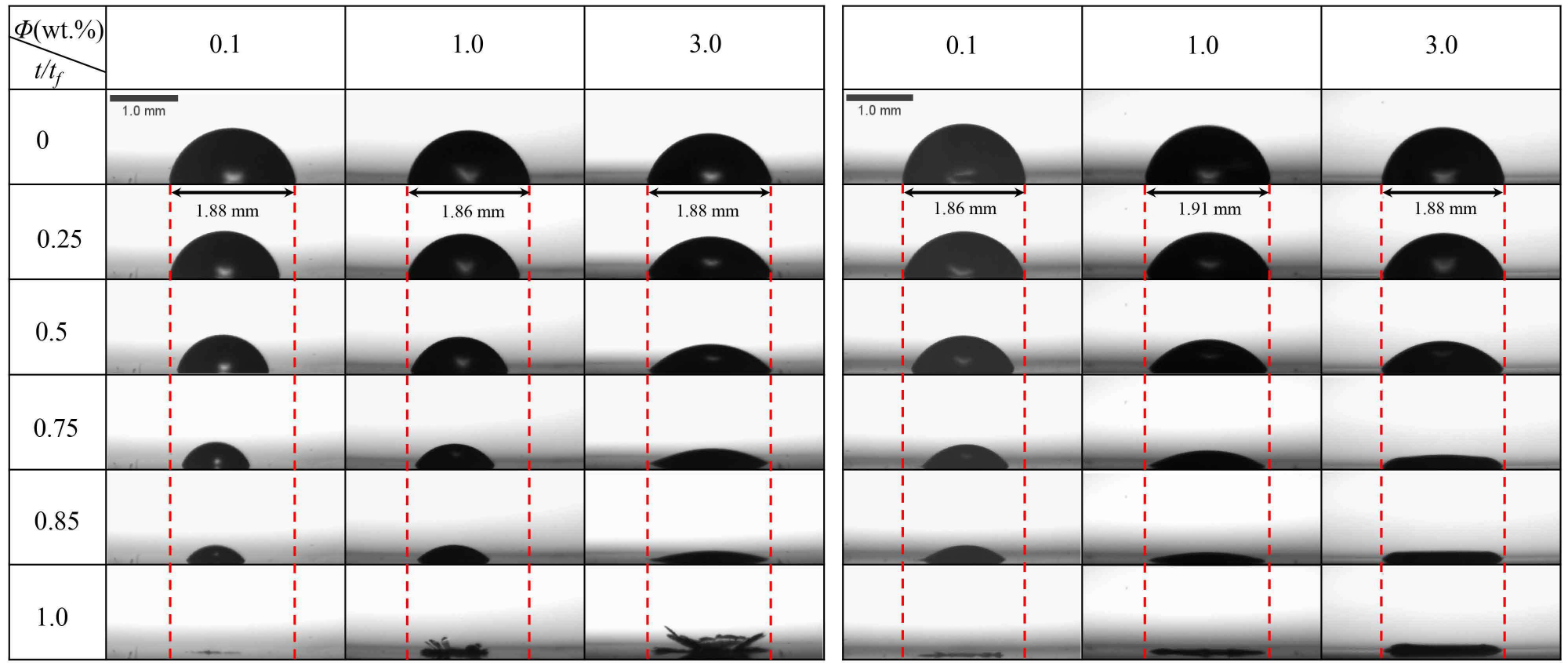}
\caption{Temporal evolution (side view) of (a) Ludox TM50 and (b) Ludox CL30 sessile droplets at different concentrations $\phi$ (0.1 wt.\%, 1.0 wt.\%, and 3.0 wt.\%) on a polystyrene substrate.} 
\label{fig:fig5}
\end{figure}

\begin{figure}
\centering
\hspace{0.6cm}{\large (a)} \hspace{4.5cm}{\large (b)} \\
\includegraphics[width=0.95\textwidth]{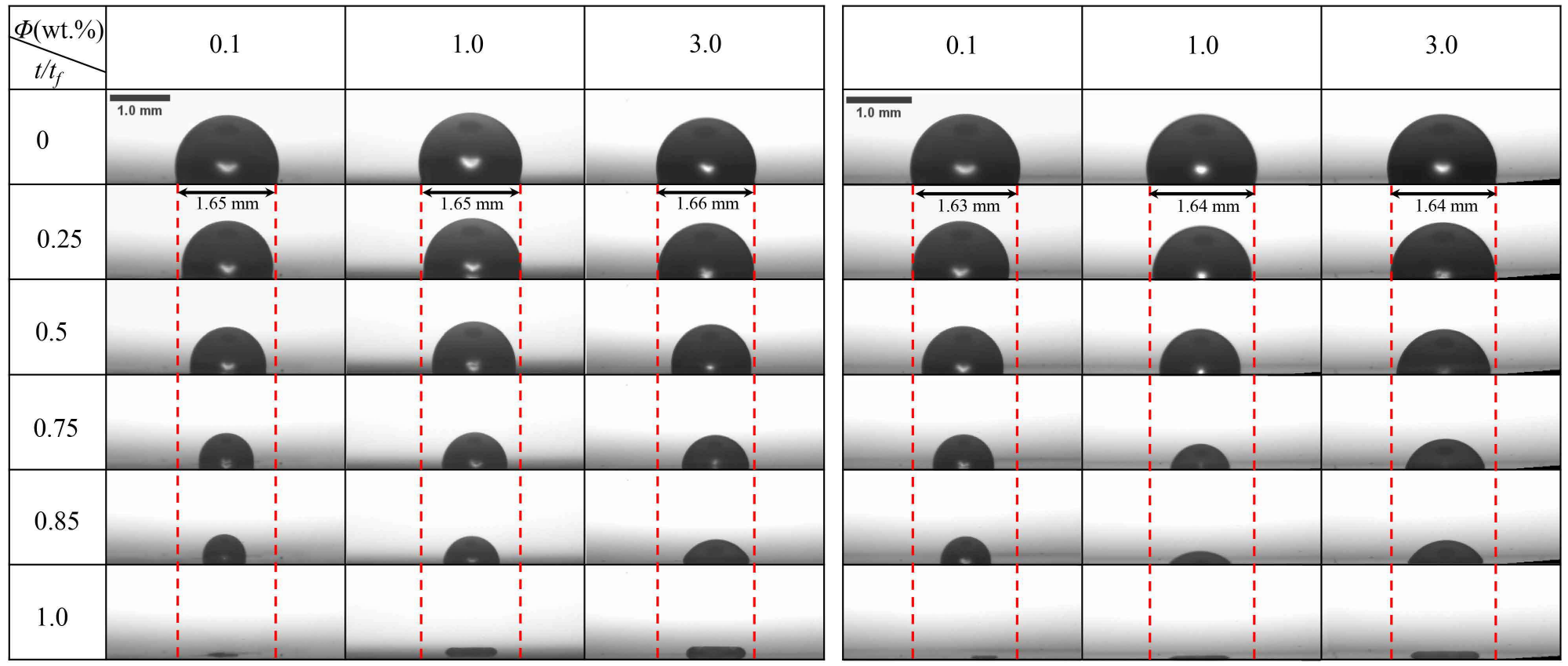}
\caption{Temporal evolution (side view) of (a) Ludox TM50 and (b) Ludox CL30 sessile droplets at different concentrations $\phi$ (0.1 wt.\%, 1.0 wt.\%, and 3.0 wt.\%) on a PTFE substrate.} 
\label{fig:fig6}
\end{figure}

Figures~\ref{fig:fig4}, \ref{fig:fig5}, and \ref{fig:fig6} illustrate the temporal evolution of Ludox TM50 and CL30 droplets at different concentrations on glass, polystyrene, and PTFE substrates, respectively. Side-view shadowgraphy images, captured at various normalized evaporation times ($t/t_f$, where $t_f$ is the total droplet lifetime), depict the drying dynamics of the nanoparticle-laden droplets. In each figure, panels (a) and (b) correspond to Ludox TM50 and CL30, respectively.

\begin{figure}
\centering
\hspace{0.5cm}{\large (a)} \hspace{4.5cm}{\large (b)} \\
\includegraphics[width=0.4\textwidth]{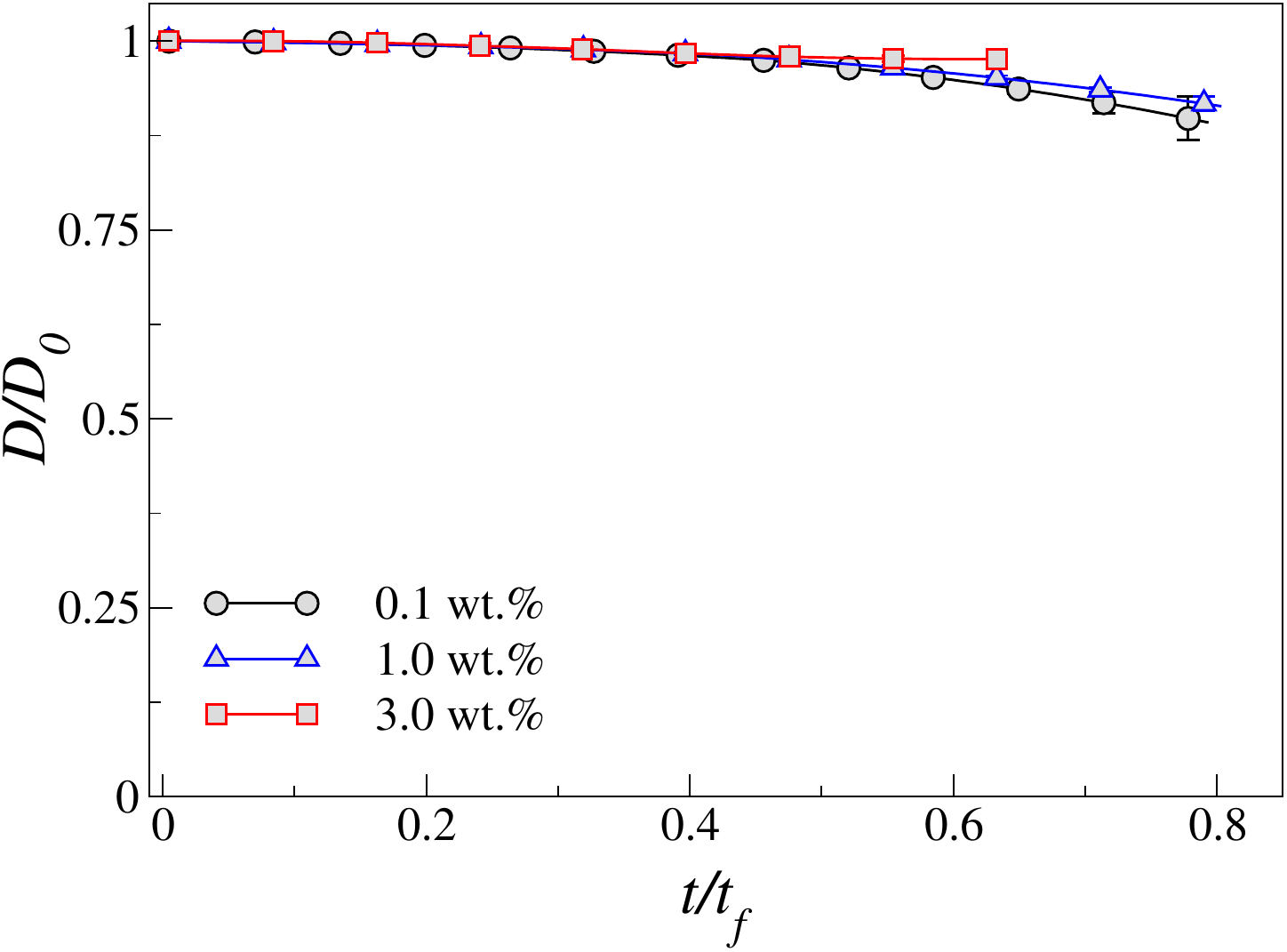} \hspace{2mm} \includegraphics[width=0.4\textwidth]{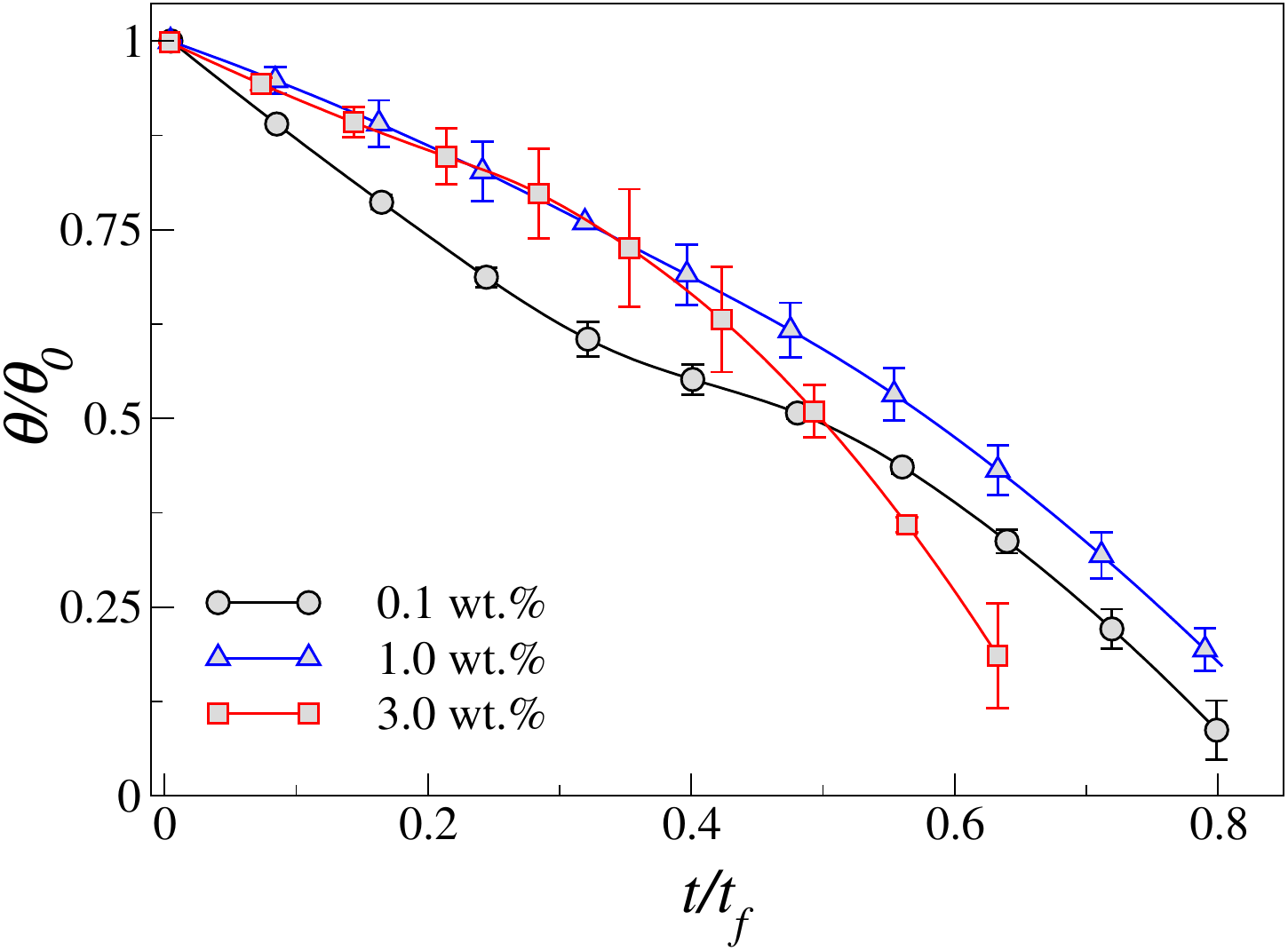}\\
\hspace{0.5cm}{\large (c)} \hspace{4.5cm}{\large (d)} \\
\includegraphics[width=0.4\textwidth]{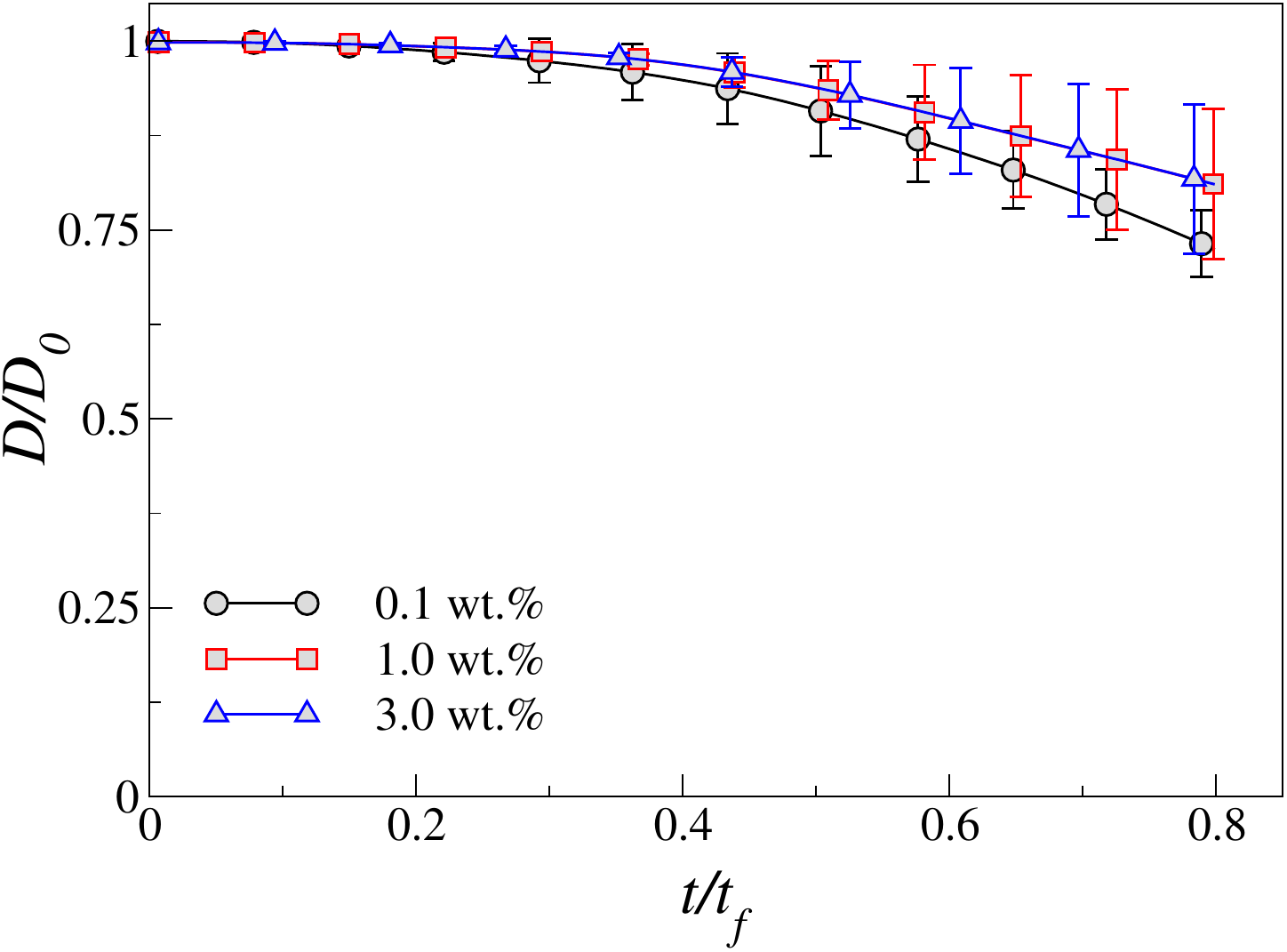} \hspace{2mm} \includegraphics[width=0.4\textwidth]{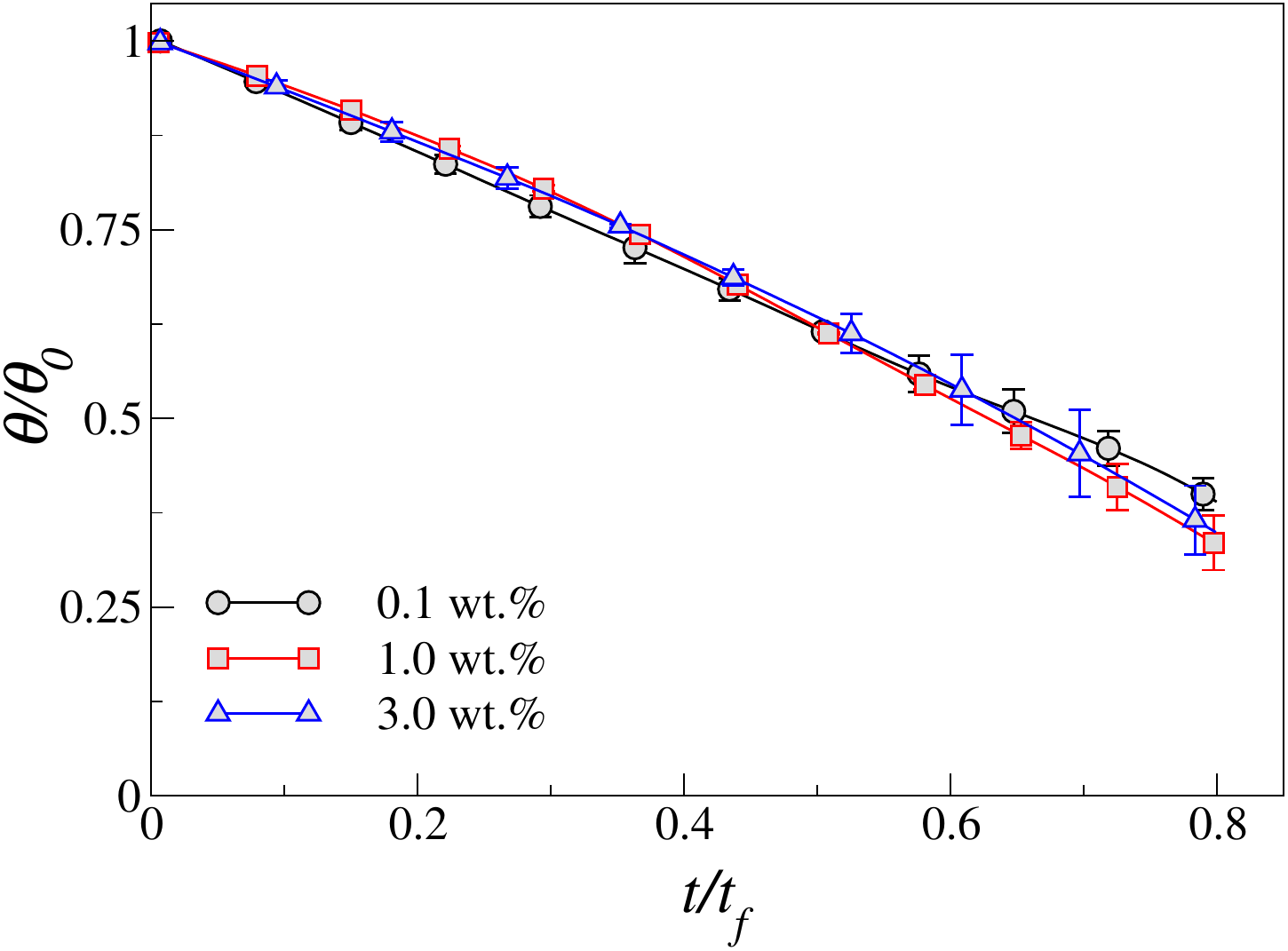}\\
\hspace{0.5cm}{\large (e)} \hspace{4.5cm}{\large (f)} \\
\includegraphics[width=0.4\textwidth]{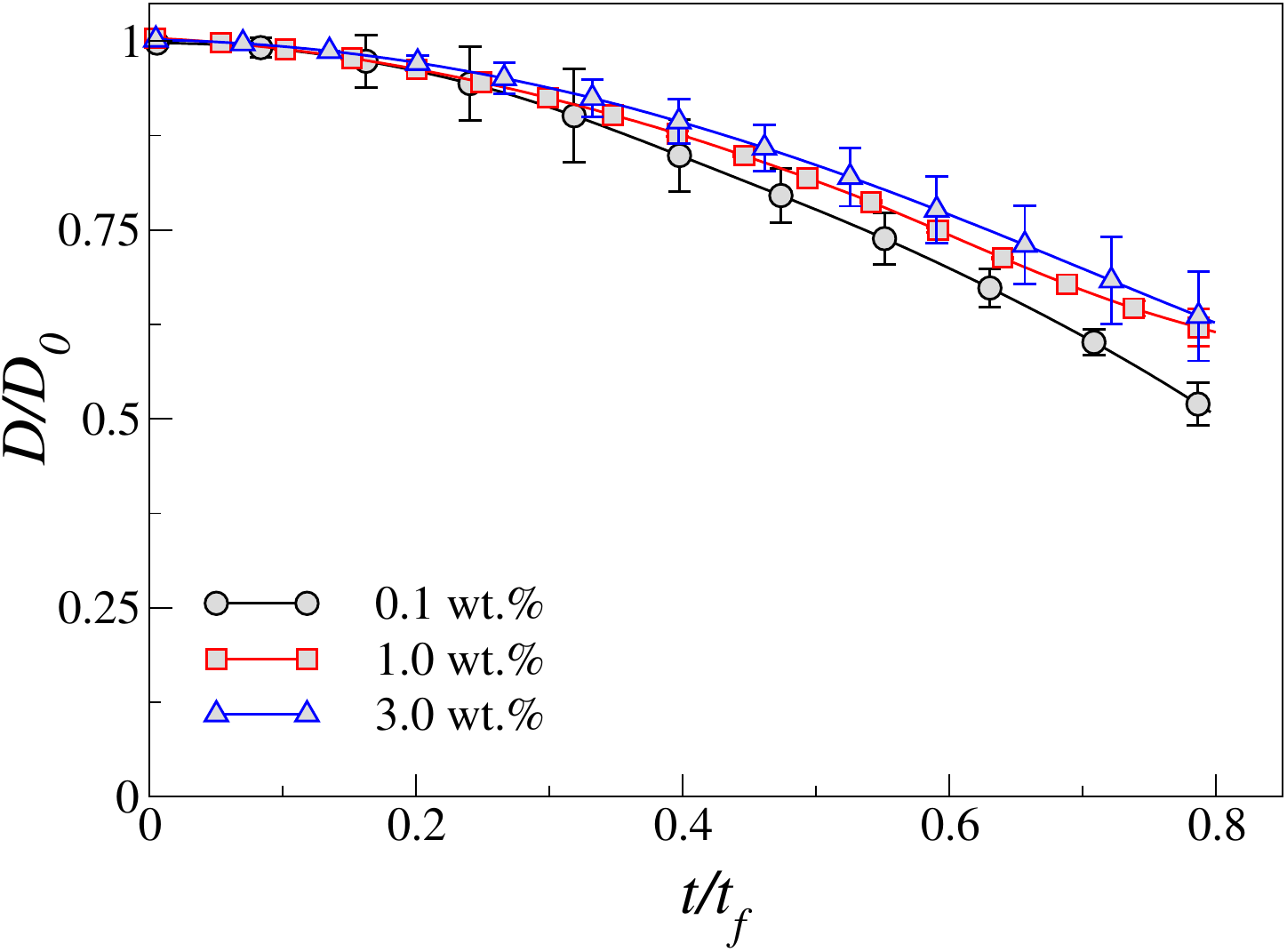} \hspace{2mm} \includegraphics[width=0.4\textwidth]{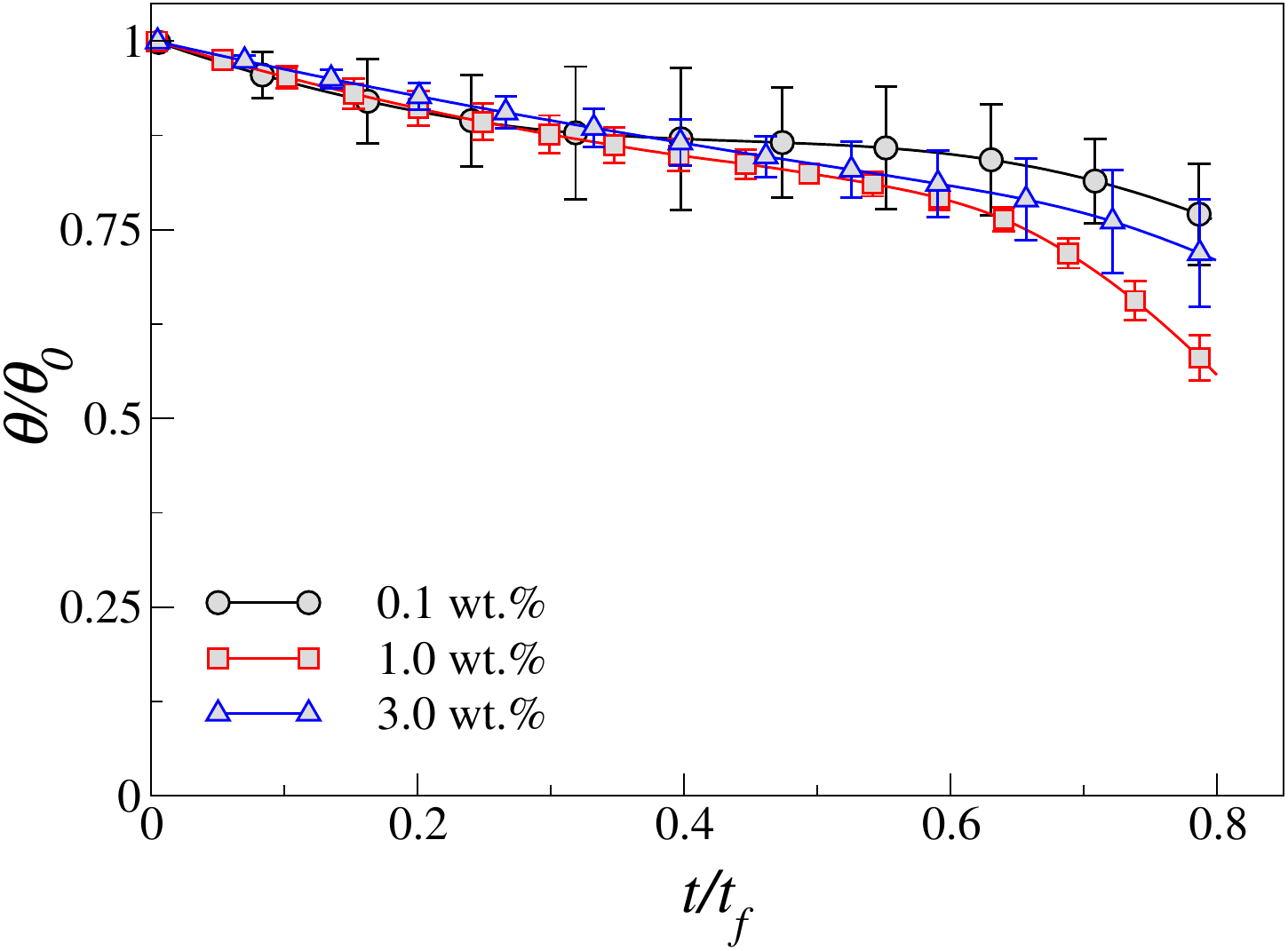}\\
\caption{Variation of normalized wetting diameter ($D/D_0$) and normalized contact angle ($\theta/\theta_0$) with normalized time ($t/t_f$) for Ludox TM50 droplets at different concentrations ($\phi$ = 0.1 wt.\%, 1.0 wt.\%, and 3.0 wt.\%) on (a, b) glass, (c, d) polystyrene, and (e, f) PTFE substrates, respectively.}
\label{fig:fig7}
\end{figure}

The evaporation dynamics of sessile droplets containing Ludox TM50 nanoparticles at various concentrations are presented in Fig. \ref{fig:fig7}(a–b), (c–d), and (e–f), which show the temporal evolution of the normalized wetting diameter ($D/D_0$) and contact angle ($\theta/\theta_0$) with respect to normalized time ($t/t_f$) on three different substrates, namely glass, polystyrene, and PTFE, respectively. It can be seen in Fig. \ref{fig:fig7}(a, b) for the glass substrate that the wetting diameter remains nearly constant over most of the droplet lifetime, indicating a constant contact radius (CCR) evaporation mode. Consequently, the contact angle exhibits a continuous and monotonic decrease. Visual confirmation of these behaviors is provided by the side-view images in Fig.~\ref{fig:fig4}(a), while Supplementary Fig. S1(a–f) presents the corresponding droplet profiles across concentrations ranging from 0.1 wt.\% to 5.0 wt.\%. The observations in Fig. \ref{fig:fig7}(a, b) are further supported by Supplementary Fig. S2(a, b), which shows a steady decline in normalized height ($h/h_0$) and volume ($V/V_0$) over time. Due to depinning of the contact line accompanied by a higher deposit thickness (Supplementary Fig. S3) for Ludox TM50 3.0 wt.\% cases, the identification of the contact line using image processing techniques were challenging. Thus the variations of the parameters are shown until $t/t_f = 0.65$.  

\begin{figure}
\centering
\hspace{0.5cm}{\large (a)} \hspace{4.5cm}{\large (b)} \\
\includegraphics[width=0.4\textwidth]{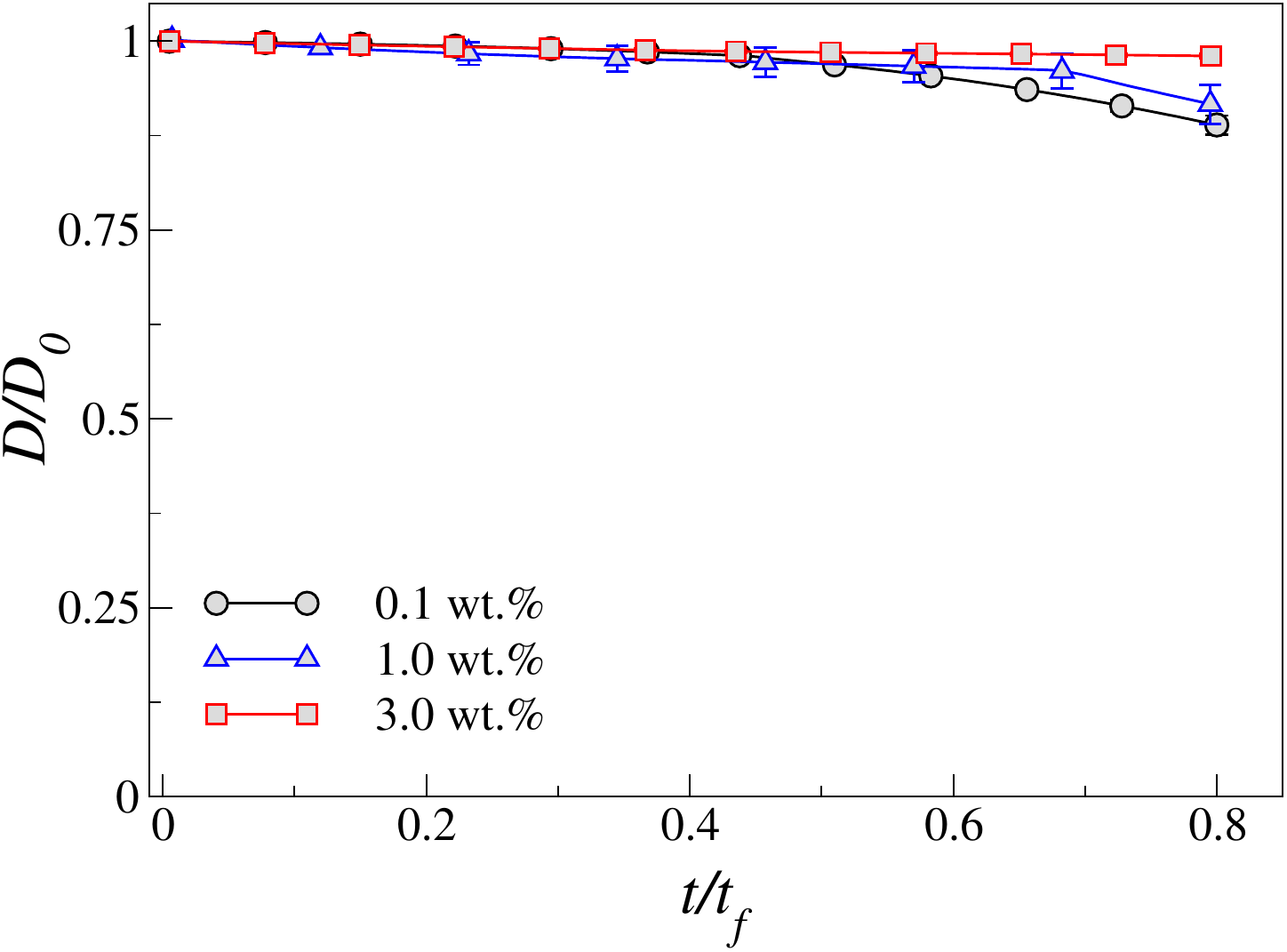} \hspace{2mm} \includegraphics[width=0.4\textwidth]{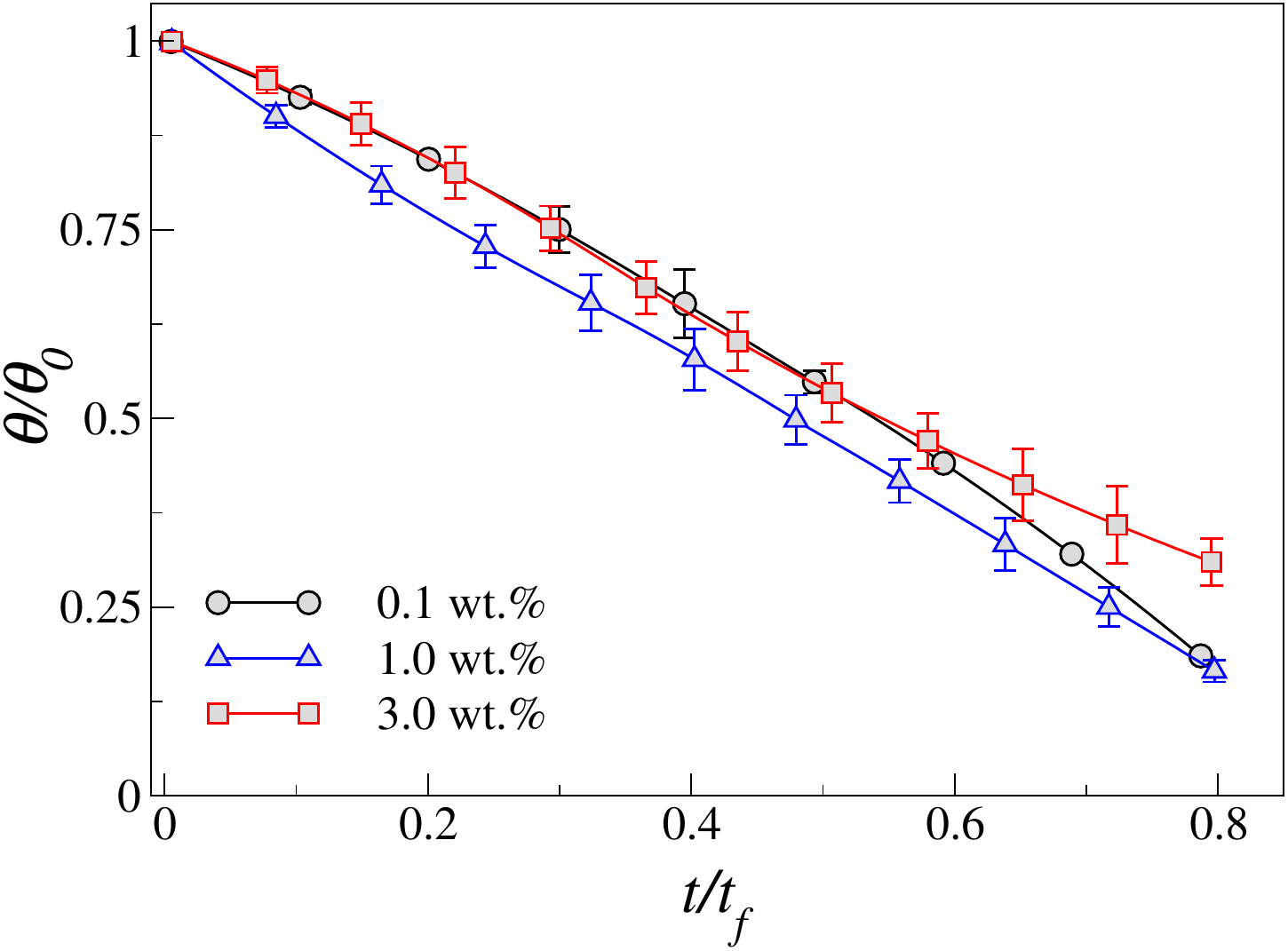}\\
\hspace{0.5cm}{\large (c)} \hspace{4.5cm}{\large (d)} \\
\includegraphics[width=0.4\textwidth]{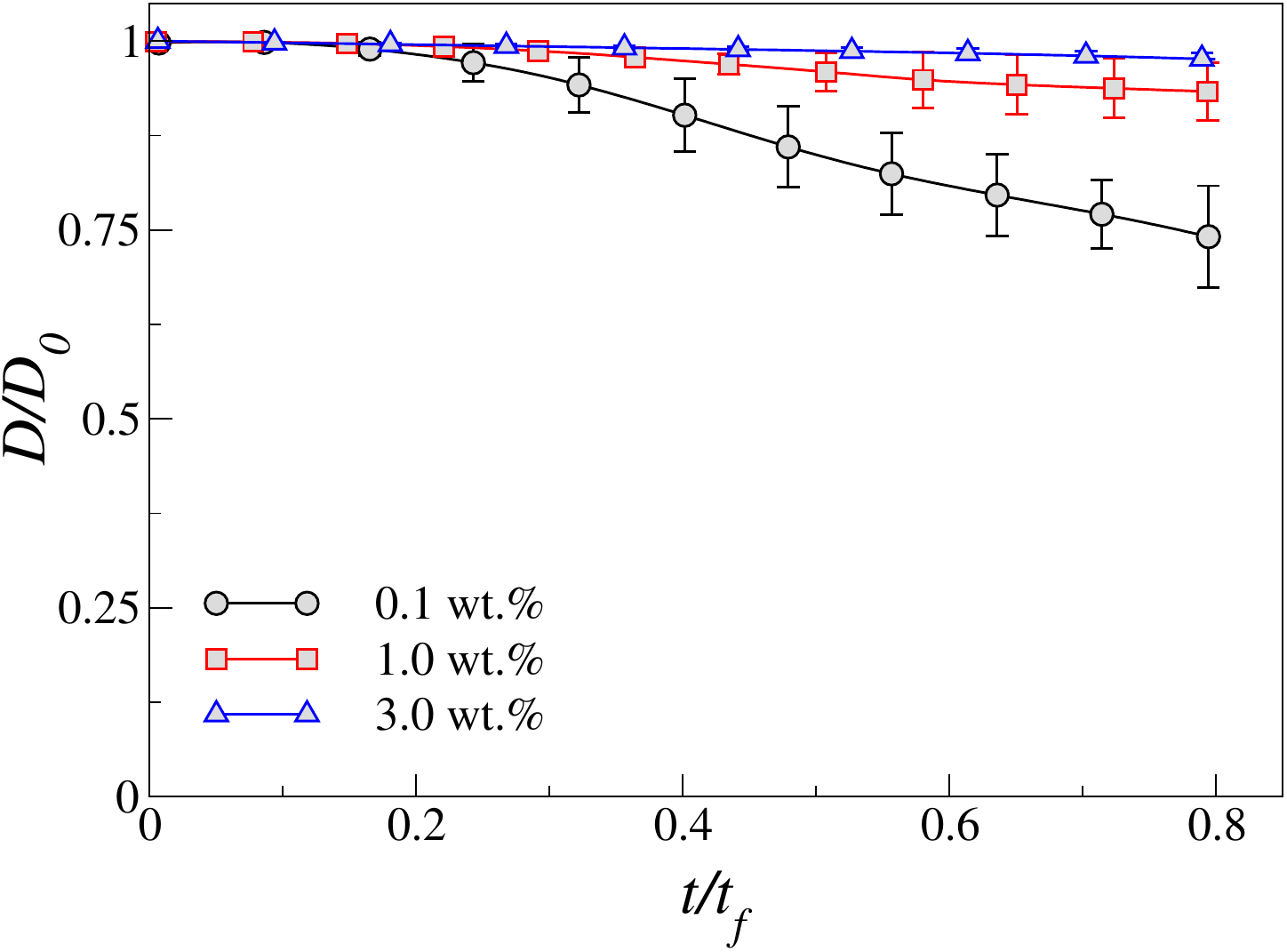} \hspace{2mm} \includegraphics[width=0.4\textwidth]{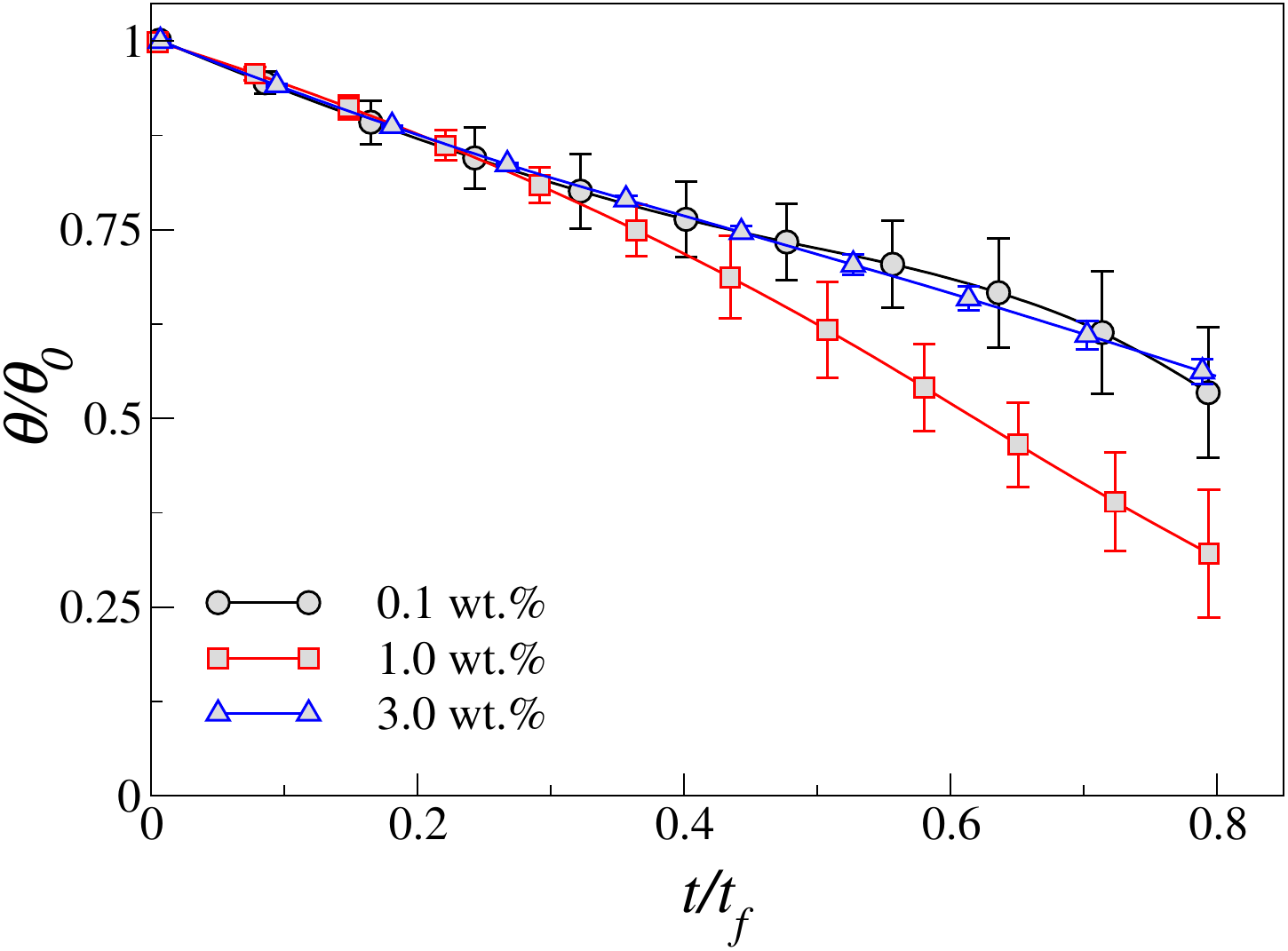}\\
\hspace{0.5cm}{\large (e)} \hspace{4.5cm}{\large (f)} \\
\includegraphics[width=0.4\textwidth]{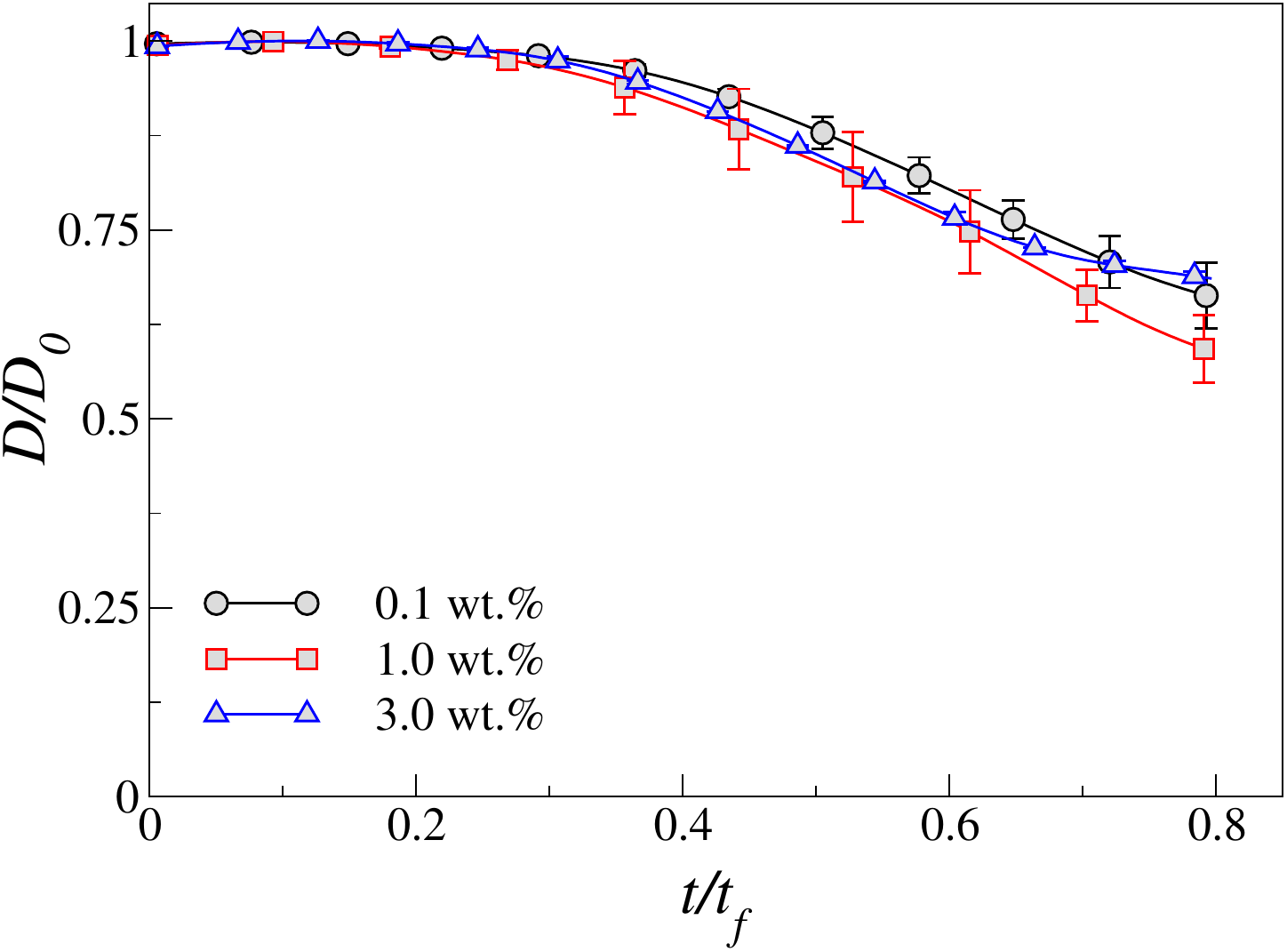} \hspace{2mm} \includegraphics[width=0.4\textwidth]{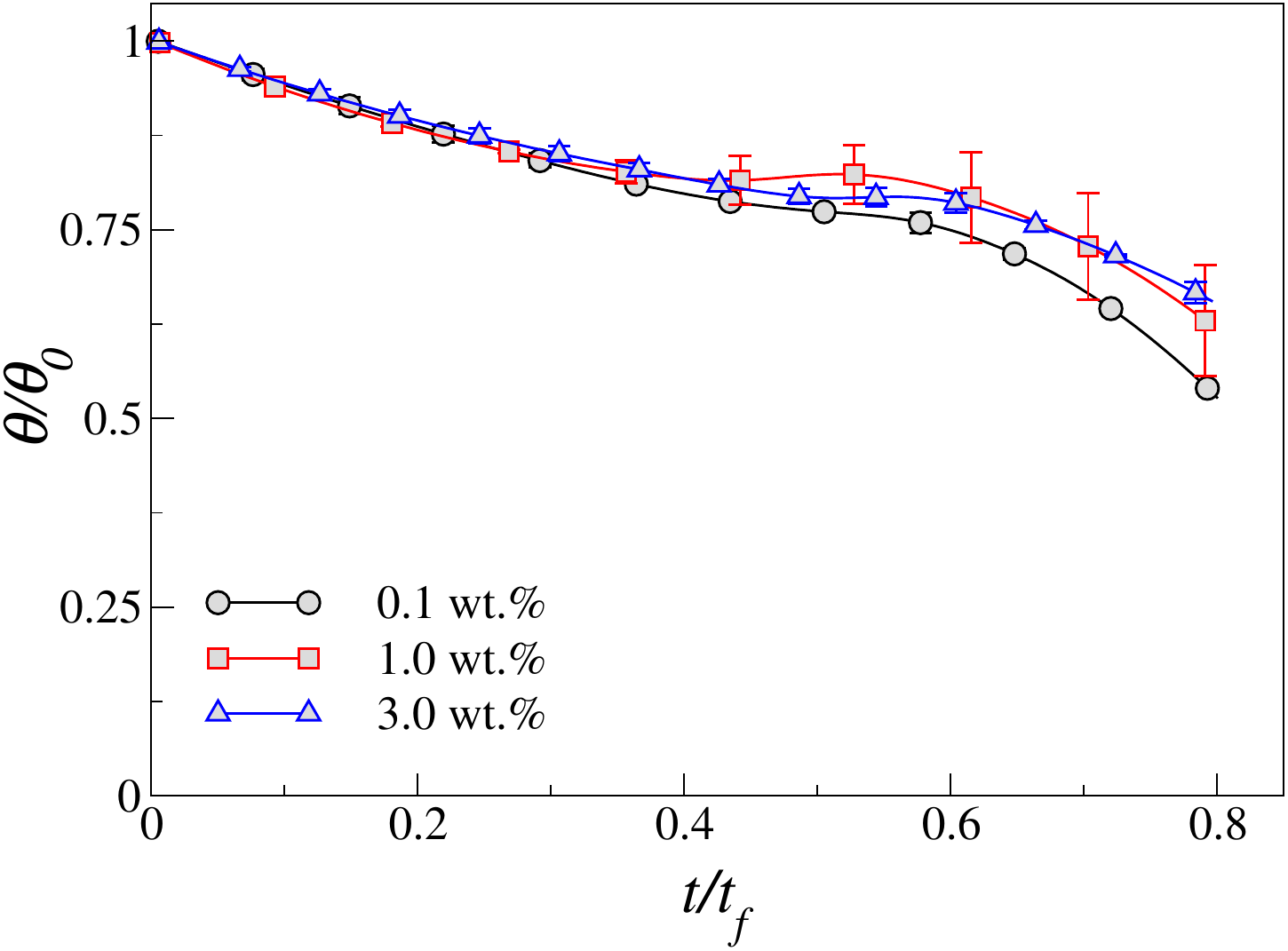}\\
\caption{Variation of normalized wetting diameter ($D/D_0$) and normalized contact angle ($\theta/\theta_0$) with normalized time ($t/t_f$) for  Ludox CL30 droplets at different concentrations ($\phi$ = 0.1 wt.\%, 1.0 wt.\%, and 3.0 wt.\%) on (a, b) glass, (c, d) polystyrene, and (e, f) PTFE substrates, respectively.}
\label{fig:fig8}
\end{figure}

On a polystyrene substrate, it can be observed in Fig. \ref{fig:fig7}(c-d) that the droplet initially exhibits a CCR mode for approximately 30\% of its lifetime, after which the contact diameter decreases. Throughout the drying process, the contact angle decreases monotonically (Fig. \ref{fig:fig7}d), a trend largely unaffected by varying concentrations of the nanoparticles (Ludox TM50). The corresponding reductions in droplet height and volume, as shown in Supplementary Fig. S2(c, d), are further corroborated by the evolution of the droplet shape captured in the side-view images in Fig. \ref{fig:fig5}(a). 

Figure~\ref{fig:fig7}(e, f) presents the evaporation kinetics of Ludox TM50 nanoparticle-laden droplets on the PTFE substrate. Unlike the evaporation behavior observed on the polystyrene substrate, the droplet on the PTFE substrate shows a continuous decrease in wetting diameter throughout the evaporation process. During the interval $0.3 \lessapprox t/t_f \lessapprox 0.6$, the contact angle remains nearly constant, indicating a prolonged constant contact angle (CCA) mode. This behavior is further supported by the normalized height and volume variations shown in Supplementary Fig. S2(e, f), along with the droplet profiles presented in Supplementary Fig. S3(a–c). Furthermore, the side-view images in Fig. \ref{fig:fig6}(a) also confirm this progression. In particular, the diameter of the final deposit increases with the concentration of nanoparticles, as evidenced by the patterns in the right panel of Fig. \ref{fig:fig2}, which is due to the particle concentration-dependent pinning of the droplet.

The evaporation characteristics of a Ludox CL30 droplet on glass are shown in Fig. \ref{fig:fig8}(a, b) for concentrations of 0.1 wt.\%, 1.0 wt.\%, and 3.0 wt.\%. All droplets follow a predominantly CCR mode, especially at higher concentrations where pinning persists until $t/t_f \approx 0.8$. For lower concentrations (0.1 wt.\% and 1.0 wt.\%), pinning lasts until $t/t_f \approx 0.6$. The accompanying decline in contact angle, height, and volume of the droplet is shown in Supplementary Fig. S4(a, b). These observations are further supported by the side-view snapshots shown in Fig. \ref{fig:fig4}(b) and droplet profiles depicted in Supplementary Fig. S1(e–l), covering a range of concentrations from 0.1 wt.\% to 5.0 wt.\%. Fig. \ref{fig:fig8}(c, d) demonstrates the evaporation behavior of a Ludox CL30 droplet on a polystyrene substrate, which also exhibits CCR-dominated evaporation regardless of concentration. Similar to the behavior on glass, the contact angle, height, and volume of the droplet on a polystyrene substrate decrease steadily throughout the drying process, as shown in Supplementary Fig. S4(c, d). Side-view images in Fig. \ref{fig:fig5}(b) visually capture these transitions. On the PTFE substrate (Fig. \ref{fig:fig8}(e, f)), the droplet initially follows a CCR mode ($0 \lessapprox {t/t_f} \lessapprox 0.3$), then transitions to a CCA mode ($0.3 \lessapprox t/t_f \lessapprox 0.6$), and finally exhibits a mixed mode during the later stages. The reduction in contact angle is accompanied by a steady decrease in height and volume as shown in Supplementary Fig. S4(e, f). This observation is also supported by the droplet profiles shown in Supplementary Fig. S3(d–f) and side-view images in Fig. \ref{fig:fig6}(b). \ks{Comparing the evaporation dynamics of Ludox CL30 and TM50 on a PTFE substrate, we observe that the pinning duration for Ludox CL30 particles ($t/t_f = 0.3$) is slightly longer than that for Ludox TM50 particles ($t/t_f = 0.2$). This difference may be attributed to attractive interactions between the oppositely charged CL30 particles and the PTFE substrate, which lead to marginally stronger contact-line pinning. However, this difference does not significantly influence the overall contact-line dynamics or the resulting deposition patterns.}

\subsection{Analysis of cracks and delamination}
\begin{figure}
\centering
\includegraphics[width=0.8\textwidth]{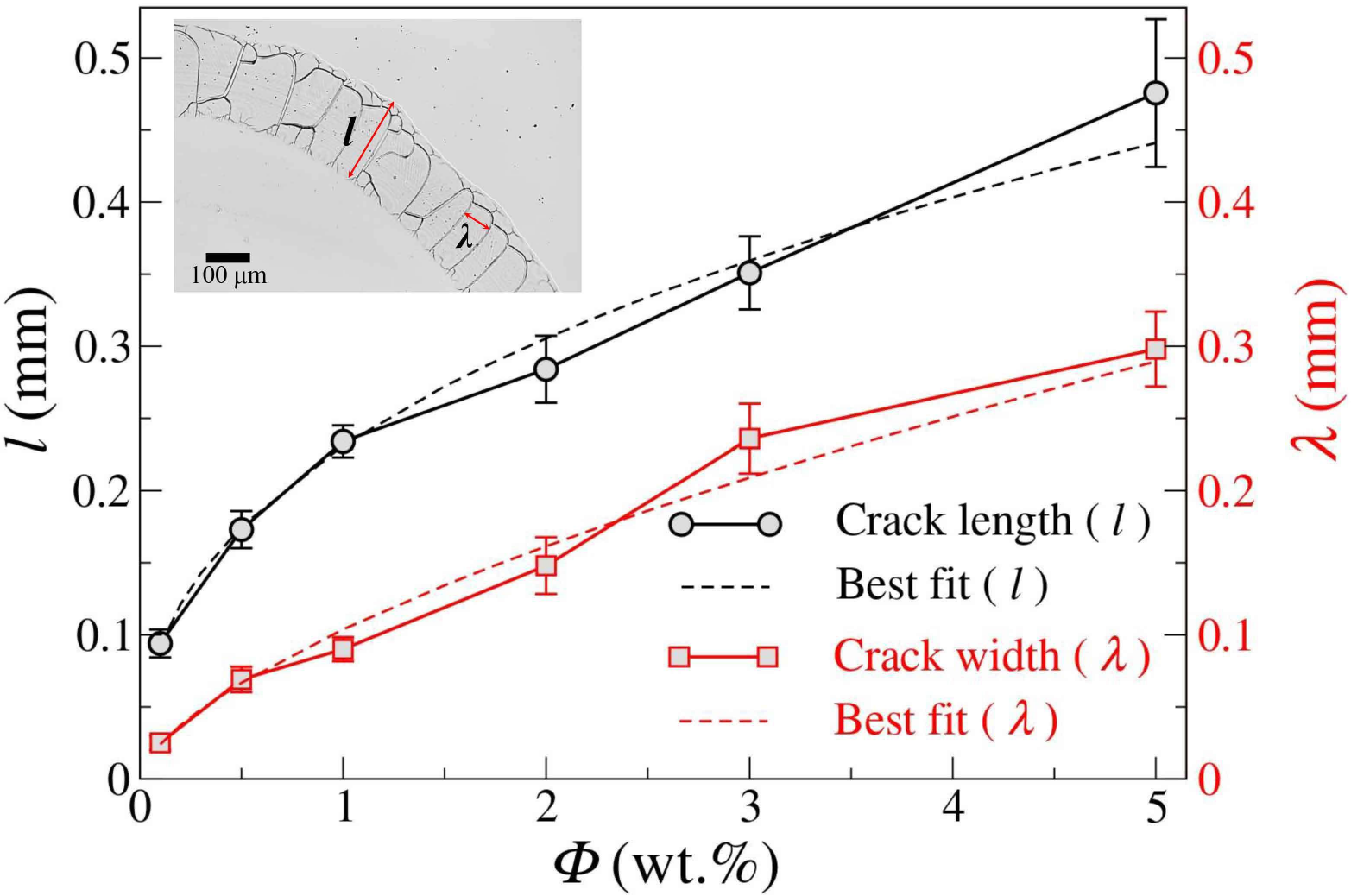}
\caption{Analysis of crack spacing and crack length in Ludox TM50 deposits on a glass substrate for concentrations ranging from $\phi = 0.1$ wt.\% to 5.0 wt.\%.} 
\label{fig:fig9}
\end{figure}

During the course of drying, as the solvent evaporates, the particles consolidate into a semi-solid particulate film, with solidification initiating at the droplet edge and progressing inward. Once the film reaches a critical thickness, mechanical instabilities such as cracking, warping, or delamination begin to appear at the periphery and propagate toward the center. A representative result showing this phenomenon is depicted in Supplementary Fig. S5. This section examines the influence of Ludox TM50 particle concentration on the onset of cracking and its characteristic features, including crack propagation, length ($l$), and spacing ($\lambda$), through analysis of the drying kinetics. The crack spacing ($\lambda$) and crack length ($l$) are analyzed using DinoCapture 2.0 (version 1.5.48.B), where the distance between two cracks is measured by drawing a line between them using the built-in tools of the software. Similarly, the line drawn along the crack from the edge towards the interior region of the deposit up to where the crack propagated gives the length of the crack (inset in Fig. \ref{fig:fig9}). Finally, at the tail end of the drying process, the particulate film undergoes progressive edge bending, resulting in symmetric upward warping or delamination that persists until the particulate film is completely dried and undergoes fragmentation. Delamination is observed beyond a critical concentration. To characterize warping or delamination, we track the strips from their initial detachment from the substrate to the end of evaporation, and analyse them using the ImageJ software. 

\begin{figure}
\centering
\includegraphics[width=0.8\textwidth]{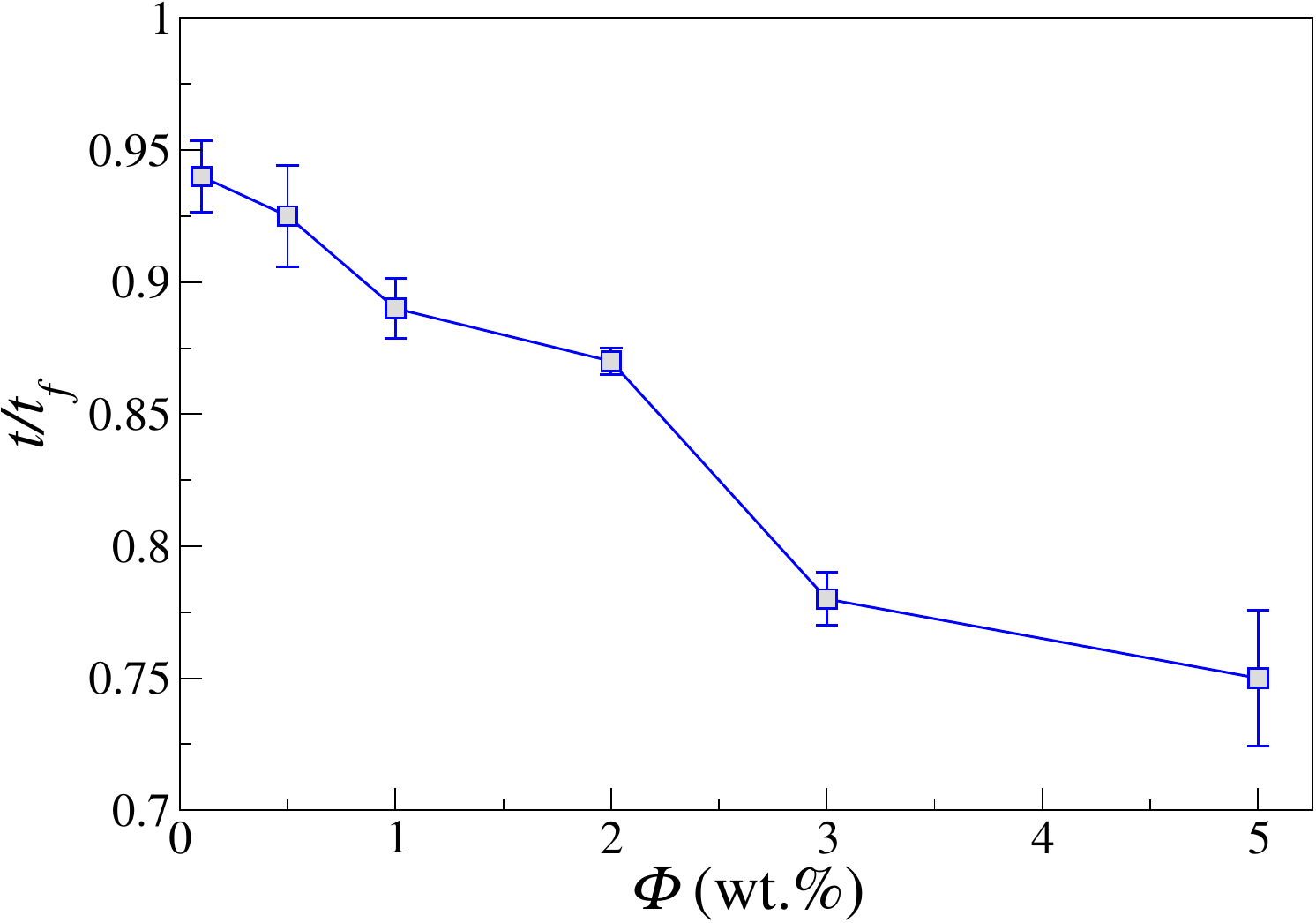}
\caption{Analysis of the onset of cracks in Ludox TM50 deposits as a function of concentration ($\phi$) on a glass substrate.} 
\label{fig:fig10}
\end{figure}

\begin{figure}
\centering
\includegraphics[width=0.8\textwidth]{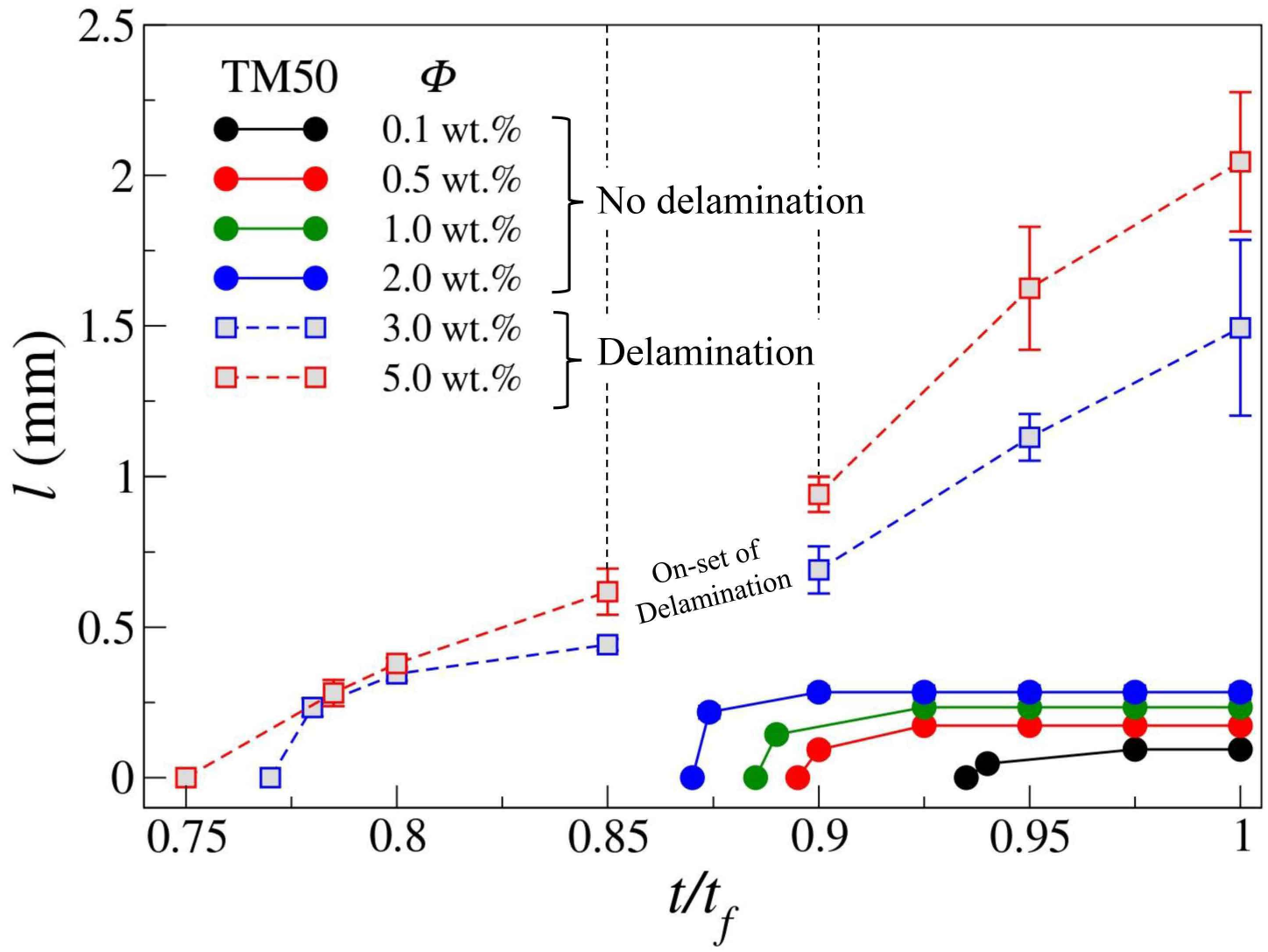}
\caption{Temporal evolution of crack length in Ludox TM50 deposits at varying concentrations $\phi$ (0.1 wt.\% – 5.0 wt.\%) on a glass substrate.} 
\label{fig:fig11}
\end{figure}

The influence of Ludox TM50 concentration on the crack morphology of the final dried deposits is systematically investigated. The average crack spacing or crack width ($\lambda$) increases with increasing Ludox TM50 concentration, ranging from 0.1 wt.\% to 5.0 wt.\%, as seen in Fig. \ref{fig:fig9}. The relationship between crack spacing and particle concentration follows a power-law dependence, expressed as $\lambda = 0.1\phi^{0.63}$. Similarly, the average crack length ($l$) shows a positive correlation with the concentration of Ludox TM50, also following the trend of the power law given by $l = 0.23\phi^{0.4}$. These relationships suggest that higher colloidal concentrations lead to the formation of wider and longer cracks in the dried films, likely due to increased stress accumulation and redistribution during the drying process. We also find that the onset of cracking largely depends on the initial particle concentration. As shown in Fig. \ref{fig:fig10} at lower particle concentration ($\phi$=0.1 wt.\%) the first crack form at $t\sim0.93t_f$ whereas, at highest particle concentration ($\phi$=5 wt.\%) first crack appears at $t\sim0.75t_f$. It has also been observed that for the concentration of Ludox TM50 from 0.1 wt.\% to 2.0 wt.\%, the cracks that originate limited to the annular ring deposit and do not propagate much as drying progress. This is due to the fact that for $\phi \le $2.0 wt.\% most of the particles are already deposited at the annular ring and overcomes the critical cracking thickness \citep{tirumkudulu2005cracking} in that region whereas the central region remains deprived of particles . To show this phenomenon, we have plotted the crack lengths starting from the onset of cracking to the end of drying for various concentrations of Ludox TM50 in Fig. \ref{fig:fig11}. Consequently, at particle concentration $\phi\geq$ 3.0 wt.\% the cracks initiate at the drop edge, and propagates till the center of the deposit which further promotes delamination.

\begin{figure}
\centering
\includegraphics[width=0.8\textwidth]{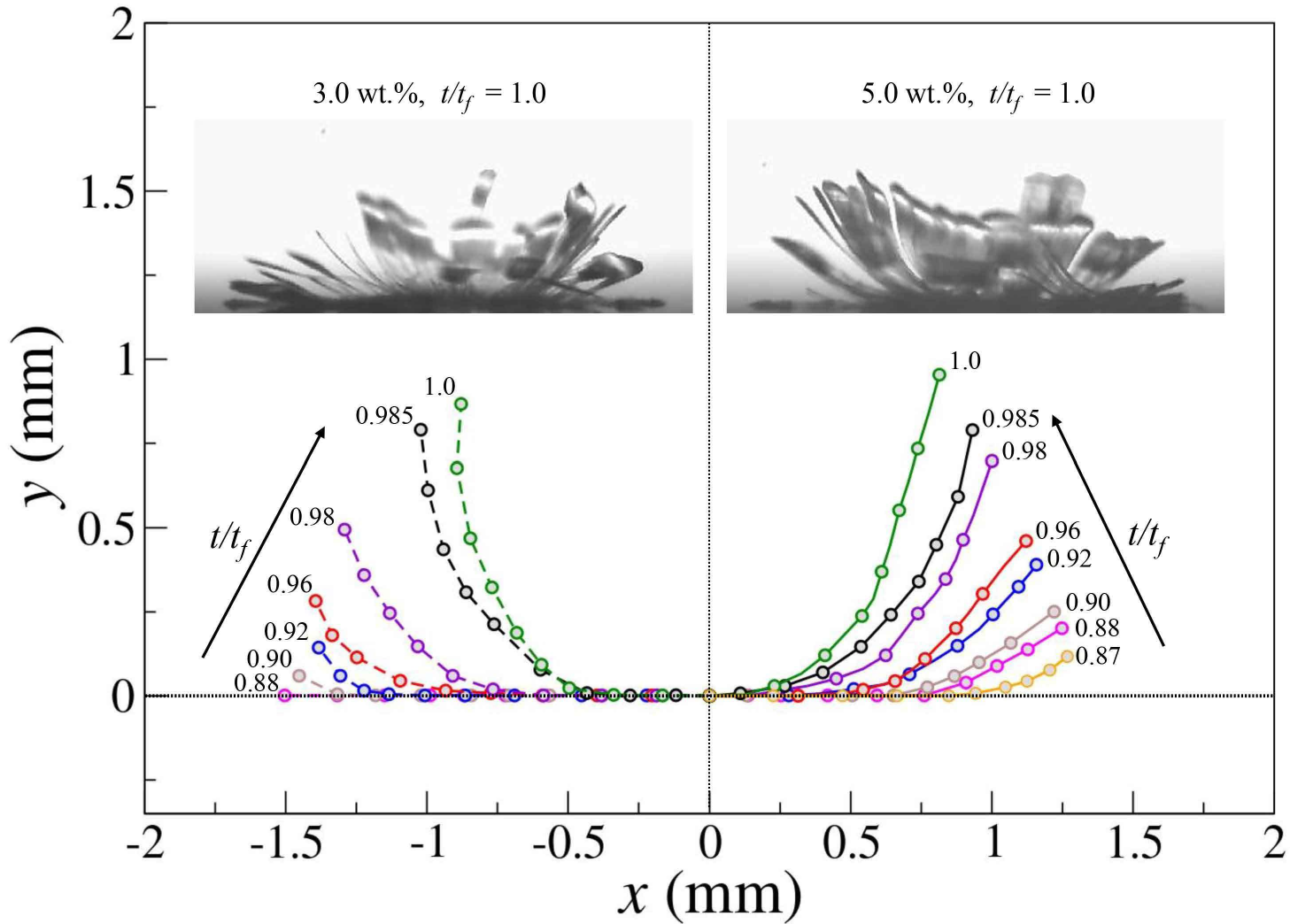}
\caption{Temporal evolution of crack delamination in Ludox TM50 deposits at concentrations of 3.0 wt.\% (left side) and 5.0 wt.\% (right side), indicated by dashed and solid lines, respectively.} 
\label{fig:fig12}
\end{figure}

\begin{figure}
\centering
\includegraphics[width=0.8\textwidth]{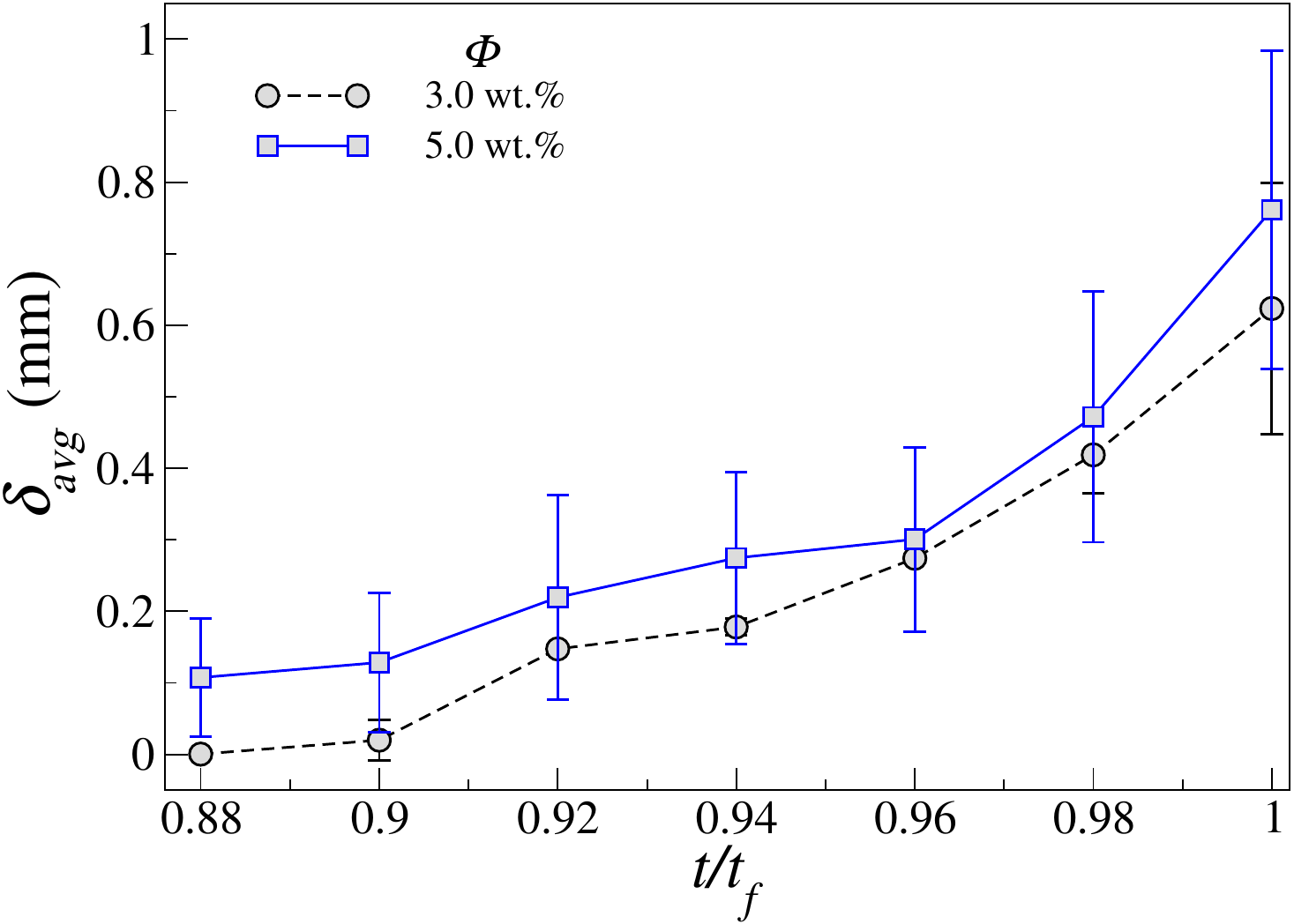}
\caption{Temporal evolution of the extent of delamination in Ludox TM50 deposits at $\phi=3.0$ wt.\% and $\phi=5.0$ wt.\%.} 
\label{fig:fig13}
\end{figure}

Next, to understand the kinetics of delamination, we examine Ludox TM50 at concentrations of 3.0 wt.\% and 5.0 wt.\%, analyzing time-lapse images of the colloidal deposit as it detaches from the substrate in both cases. We track the coordinates of a specific strip or crack in the deposit at various time intervals to quantify its deflection during delamination. The extent of delamination is determined by measuring the deflection height from the substrate over time. Fig. \ref{fig:fig12} shows the temporal evolution of the deflection of delaminating strip for 3.0 wt.$\%$ (left side of the figure) and 5.0 wt.$\%$ (right side of the figure). Fig. \ref{fig:fig13} depicts the extent of delamination ($\delta_{avg}$) at different intervals of time during the evaporation of Ludox TM50. Figs. \ref{fig:fig12} and \ref{fig:fig13} demonstrate that a higher particle concentration promotes faster delamination with a greater extent. Further, it can also be seen that delamination is more pronounced on the glass substrate than on the polystyrene surface. Supplementary Fig. S6 presents a comparison of the temporal evolution of delamination for 3.0 wt.\% Ludox TM50 deposits on glass and polystyrene substrates. This enhanced delamination on glass is attributed to its higher wettability, which facilitates greater initial spreading of the droplet. The increased spreading results in a larger deposit area, leading to longer crack lengths, and consequently, a greater extent of delamination during the drying process. 

We also found that on glass substrates, delamination is observed only at higher concentrations ($\ge 3$ wt.\%) due to the greater thickness of the deposited film, which generates a larger bending moment that promotes delamination. In contrast, on polystyrene substrates, the initial droplet spread is smaller, leading to a thicker deposit at the same concentration. As a result, delamination occurs at lower initial concentrations (1 wt.\%) on the polystyrene substrate compared to the glass substrate, as shown in Figs. \ref{fig:fig4} and \ref{fig:fig5}, respectively.

\begin{figure}
\centering
\includegraphics[width=0.48\textwidth]{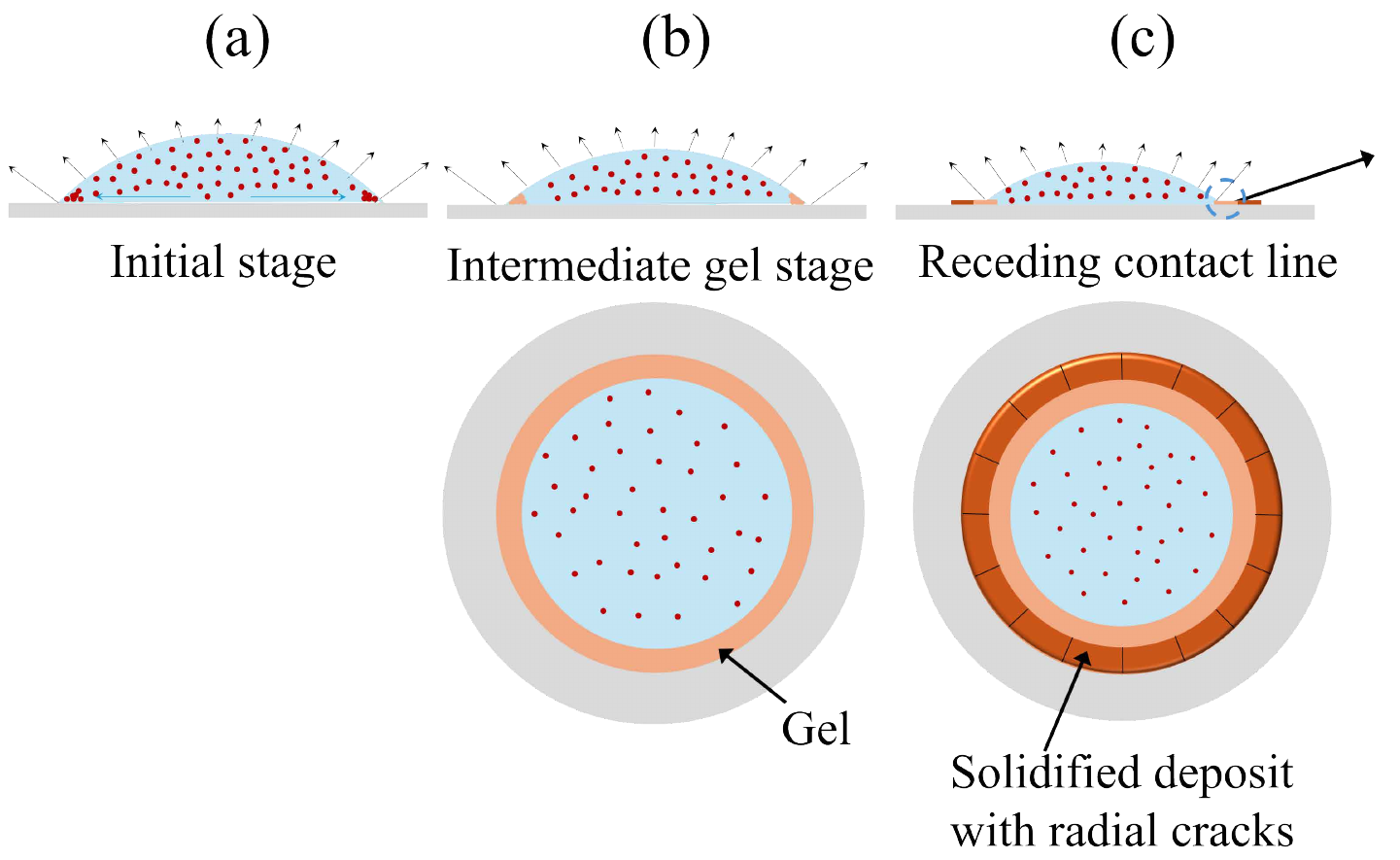} 
\includegraphics[width=0.48\textwidth]{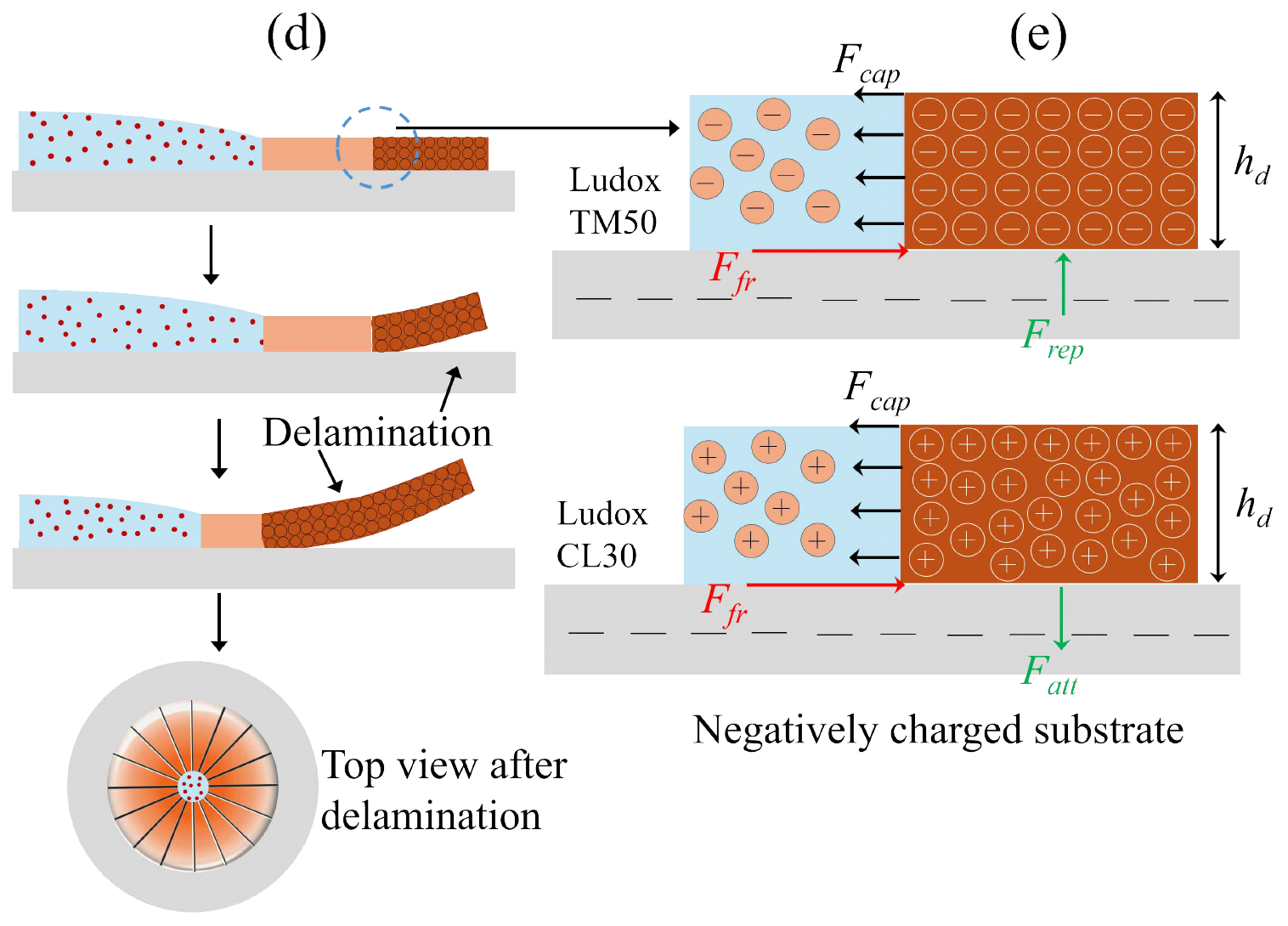} \\
\caption{A mechanism explaining the influence of capillary forces, interfacial interactions, and attractive or repulsive particle–substrate forces in governing the cracking and delamination behavior observed in Ludox TM50 colloidal suspensions on glass and polystyrene substrates.}
\label{fig:fig14}
\end{figure}

Finally, the underlying mechanism of crack formation and delamination is schematically illustrated in Fig. \ref{fig:fig14}. During the initial stage, as the sessile droplet evaporates, the Ludox particles are carried to the triple contact line by the radial capillary flows from the bulk of the droplet. This flow results in the accumulation of particles near the pinned triple contact line as shown in Fig. \ref{fig:fig14}(a). As evaporation progresses, colloidal particles consolidate near the contact line, leading to the formation of a gel-like state. Solidification then initiates at the droplet edge and advances inward, as shown in Fig. \ref{fig:fig14}(b). As the upper part of the deposit shrinks more rapidly than the region adhered to the substrate, unbalanced tensile stress develops, which is subsequently relieved through the formation of cracks (Fig. \ref{fig:fig14}c). As discussed earlier, the evaporation kinetics is largely unaffected by the surface charge of the particles. Thus, the formation of initial cracks is independent of the particle charge type. However, the characteristics of the resulting cracks and the subsequent delamination (Fig.\ref{fig:fig14}d) are influenced by the particle surface charge on a given substrate. Fig.\ref{fig:fig14}(e) illustrates two different scenarios considered in the present study, involving Ludox TM50 (negatively charged) and Ludox CL30 (positively charged) on a glass substrate (negatively charged). For a given film thickness ($h_d$), the transition from a gel to a solid state begins at the droplet edge and progresses inward as drying continues. In the gel region, the colloidal particles experience compressive capillary forces ($F_{cap}$), which are also transmitted to the adhered solidified region, pulling it inward. At the same time, the solidified region experiences a frictional force ($F_{fr}$) from the substrate in the direction opposite to $F_{cap}$. The particle-substrate charge interaction is also crucial. While the oppositely charged systems promote adhesion, the like-charged systems are prone to repel. When the capillary force ($F_{cap}$) significantly exceeds the frictional force ($F_{fr}$), the solidified strip begins to delaminate. This delamination extends the pre-existing radial crack further toward the center of the droplet until the entire strip detaches from the substrate (Fig. \ref{fig:fig14}(d)). In the Ludox TM50 system, the repulsive particle–substrate interaction promotes the formation of well-ordered radial cracks, followed by complete delamination of the film. In contrast, in the Ludox CL30 system, the attractive interaction between particles and the substrate leads to randomly oriented cracks and prevents delamination.

\ks{Furthermore, it is to be noted that the delamination occurs when the elastic strain energy stored in the drying film exceeds the film–substrate adhesion energy \cite{pauchard2006patterns}. In the present study, the delamination height, denoted by $\delta_{avg}$, quantifies the extent of film lifting and serves as a measurable indicator of delamination. Specifically, $\delta_{avg}$ at any instant of time is defined as the vertical distance from the substrate to the tip of a delaminating strip, as obtained from side-view CMOS camera images during the drying of the colloidal drop. For consistency, the leftmost and rightmost delaminating strips were considered for measurement, and the reported values were averaged over three independent repetitions for particle concentrations of $3$ and $5$ wt.\%. The driving force for delamination arises from tensile stresses generated during solvent evaporation and particle consolidation within the film. On glass substrates, greater initial spreading leads to the formation of a thinner and more laterally extended film, which enhances in-plane tensile stresses during drying. Consequently, the elastic strain energy accumulated in the film is higher, making delamination more pronounced once the adhesion energy threshold is exceeded. In contrast, reduced spreading on PTFE limits stress buildup and thereby suppresses delamination.}

\section{Conclusions} \label{sec:conc}

We investigate the influence of particle surface charge and substrate wettability on the evaporation dynamics, deposition morphology, and cracking/ delamination behavior of colloidal silica suspensions. By employing negatively charged Ludox TM50 and positively charged Ludox CL30 nanoparticles on substrates of varying wettability, namely glass, polystyrene, and PTFE, we observe that the evaporation mode is strongly dictated by the substrate. On glass, droplets predominantly follow a constant contact radius (CCR) mode, while on PTFE, a prolonged constant contact angle (CCA) mode is observed. These variations significantly affect the final deposit morphology, with glass substrates yielding broader spread and thicker deposits. The results reveal that particle concentration plays a crucial role in determining crack morphology. For Ludox TM50, both crack spacing and length increase with concentration, following distinct power-law relations. At low concentrations, cracks appear only in the final stages of evaporation, whereas at higher concentrations, they initiate early and propagate toward the droplet center. In contrast, Ludox CL30 droplets form irregular, randomly oriented cracks due to attractive electrostatic interactions with the negatively charged glass substrate. Delamination behavior is also found to depend on both particle concentration and substrate type. On glass, delamination occurs only at higher concentrations ($\ge 3.0$ wt.\%). However, on polystyrene substrates, where the droplet spread is lower, delamination is observed even at lower concentrations (1.0 wt.\%), highlighting the role of local thickness and interfacial stress. Temporal tracking of delamination shows that the extent of upward bending is greater for higher concentrations and more pronounced on glass substrates. A mechanistic framework is proposed to explain these phenomena, emphasising the roles of capillary-induced particle–particle and particle–substrate interactions.

\vspace{2mm}
\noindent{\bf Acknowledgement:} {K. C. S. and R. M. thank IIT Hyderabad for financial support through grants IITH/CHE/ F011/SOCH1 and SG/IITH/F305/2022-23/SG-143, respectively.}


\begin{thebibliography}{69}
\expandafter\ifx\csname natexlab\endcsname\relax\def\natexlab#1{#1}\fi
\providecommand{\url}[1]{\texttt{#1}}
\providecommand{\href}[2]{#2}
\providecommand{\path}[1]{#1}
\providecommand{\DOIprefix}{doi:}
\providecommand{\ArXivprefix}{arXiv:}
\providecommand{\URLprefix}{URL: }
\providecommand{\Pubmedprefix}{pmid:}
\providecommand{\doi}[1]{\href{http://dx.doi.org/#1}{\path{#1}}}
\providecommand{\Pubmed}[1]{\href{pmid:#1}{\path{#1}}}
\providecommand{\bibinfo}[2]{#2}
\ifx\xfnm\relax \def\xfnm[#1]{\unskip,\space#1}\fi
\bibitem[{Bamboriya and Tirumkudulu(2025)}]{bamboriya2025effective}
\bibinfo{author}{Bamboriya, O.P.}, \bibinfo{author}{Tirumkudulu, M.S.},
  \bibinfo{year}{2025}.
\newblock \bibinfo{title}{Effective modulus of particle-packing containing hard
  and soft particles}.
\newblock \bibinfo{journal}{Soft Matter} \bibinfo{volume}{21},
  \bibinfo{pages}{2986–2993}.
\bibitem[{Bhardwaj et~al.(2010)Bhardwaj, Fang, Somasundaran and
  Attinger}]{bhardwaj2010self}
\bibinfo{author}{Bhardwaj, R.}, \bibinfo{author}{Fang, X.},
  \bibinfo{author}{Somasundaran, P.}, \bibinfo{author}{Attinger, D.},
  \bibinfo{year}{2010}.
\newblock \bibinfo{title}{Self-assembly of colloidal particles from evaporating
  droplets: role of dlvo interactions and proposition of a phase diagram}.
\newblock \bibinfo{journal}{Langmuir} \bibinfo{volume}{26},
  \bibinfo{pages}{7833--7842}.
\bibitem[{Bourrianne et~al.(2021)Bourrianne, Lilin, Sint{\`e}s, N{\^\i}rca,
  McKinley and Bischofberger}]{bourrianne2021crack}
\bibinfo{author}{Bourrianne, P.}, \bibinfo{author}{Lilin, P.},
  \bibinfo{author}{Sint{\`e}s, G.}, \bibinfo{author}{N{\^\i}rca, T.},
  \bibinfo{author}{McKinley, G.H.}, \bibinfo{author}{Bischofberger, I.},
  \bibinfo{year}{2021}.
\newblock \bibinfo{title}{Crack morphologies in drying suspension drops}.
\newblock \bibinfo{journal}{Soft Matter} \bibinfo{volume}{17},
  \bibinfo{pages}{8832--8837}.
\bibitem[{Bridonneau et~al.(2020)Bridonneau, Zhao, Battaglini, Mattana,
  Th{\'e}venet, No{\"e}l, Roch{\'e}, Zrig and Carn}]{bridonneau2020self}
\bibinfo{author}{Bridonneau, N.}, \bibinfo{author}{Zhao, M.},
  \bibinfo{author}{Battaglini, N.}, \bibinfo{author}{Mattana, G.},
  \bibinfo{author}{Th{\'e}venet, V.}, \bibinfo{author}{No{\"e}l, V.},
  \bibinfo{author}{Roch{\'e}, M.}, \bibinfo{author}{Zrig, S.},
  \bibinfo{author}{Carn, F.}, \bibinfo{year}{2020}.
\newblock \bibinfo{title}{Self-assembly of nanoparticles from evaporating
  sessile droplets: fresh look into the role of particle/substrate
  interaction}.
\newblock \bibinfo{journal}{Langmuir} \bibinfo{volume}{36},
  \bibinfo{pages}{11411--11421}.
\bibitem[{Brinckmann and Stephan(2011)}]{brinckmann2011experimental}
\bibinfo{author}{Brinckmann, F.}, \bibinfo{author}{Stephan, P.},
  \bibinfo{year}{2011}.
\newblock \bibinfo{title}{Experimental investigation of the drying process of
  water-based paints used in automotive industry}.
\newblock \bibinfo{journal}{Chem. Eng. Process.: Process Intensif.}
  \bibinfo{volume}{50}, \bibinfo{pages}{489--494}.
\bibitem[{Chang and Velev(2006)}]{chang2006evaporation}
\bibinfo{author}{Chang, S.T.}, \bibinfo{author}{Velev, O.D.},
  \bibinfo{year}{2006}.
\newblock \bibinfo{title}{Evaporation-induced particle microseparations inside
  droplets floating on a chip}.
\newblock \bibinfo{journal}{Langmuir} \bibinfo{volume}{22},
  \bibinfo{pages}{1459--1468}.
\bibitem[{Chhasatia et~al.(2010)Chhasatia, Joshi and Sun}]{chhasatia2010effect}
\bibinfo{author}{Chhasatia, V.H.}, \bibinfo{author}{Joshi, A.S.},
  \bibinfo{author}{Sun, Y.}, \bibinfo{year}{2010}.
\newblock \bibinfo{title}{Effect of relative humidity on contact angle and
  particle deposition morphology of an evaporating colloidal drop}.
\newblock \bibinfo{journal}{Appl. Phys. Lett.} \bibinfo{volume}{97},
  \bibinfo{pages}{231909}.
\bibitem[{Davoust and Theisen(2013)}]{davoust2013evaporation}
\bibinfo{author}{Davoust, L.}, \bibinfo{author}{Theisen, J.},
  \bibinfo{year}{2013}.
\newblock \bibinfo{title}{Evaporation rate of drop arrays within a digital
  microfluidic system}.
\newblock \bibinfo{journal}{Sens. Actuators, B Chem.} \bibinfo{volume}{189},
  \bibinfo{pages}{157--164}.
\bibitem[{Deegan et~al.(1997)Deegan, Bakajin, Dupont, Huber, Nagel and
  Witten}]{deegan1997capillary}
\bibinfo{author}{Deegan, R.D.}, \bibinfo{author}{Bakajin, O.},
  \bibinfo{author}{Dupont, T.F.}, \bibinfo{author}{Huber, G.},
  \bibinfo{author}{Nagel, S.R.}, \bibinfo{author}{Witten, T.A.},
  \bibinfo{year}{1997}.
\newblock \bibinfo{title}{Capillary flow as the cause of ring stains from dried
  liquid drops}.
\newblock \bibinfo{journal}{Nature} \bibinfo{volume}{389},
  \bibinfo{pages}{827--829}.
\bibitem[{Deegan et~al.(2000)Deegan, Bakajin, Dupont, Huber, Nagel and
  Witten}]{deegan2000contact}
\bibinfo{author}{Deegan, R.D.}, \bibinfo{author}{Bakajin, O.},
  \bibinfo{author}{Dupont, T.F.}, \bibinfo{author}{Huber, G.},
  \bibinfo{author}{Nagel, S.R.}, \bibinfo{author}{Witten, T.A.},
  \bibinfo{year}{2000}.
\newblock \bibinfo{title}{Contact line deposits in an evaporating drop}.
\newblock \bibinfo{journal}{Phys. Rev. E.} \bibinfo{volume}{62},
  \bibinfo{pages}{756--765}.
\bibitem[{Dugyala and Basavaraj(2014)}]{dugyala2014control}
\bibinfo{author}{Dugyala, V.R.}, \bibinfo{author}{Basavaraj, M.G.},
  \bibinfo{year}{2014}.
\newblock \bibinfo{title}{Control over coffee-ring formation in evaporating
  liquid drops containing ellipsoids}.
\newblock \bibinfo{journal}{Langmuir} \bibinfo{volume}{30},
  \bibinfo{pages}{8680--8686}.
\bibitem[{Dugyala et~al.(2016)Dugyala, Lama, Satapathy and
  Basavaraj}]{dugyala2016role}
\bibinfo{author}{Dugyala, V.R.}, \bibinfo{author}{Lama, H.},
  \bibinfo{author}{Satapathy, D.K.}, \bibinfo{author}{Basavaraj, M.G.},
  \bibinfo{year}{2016}.
\newblock \bibinfo{title}{Role of particle shape anisotropy on crack formation
  in drying of colloidal suspension}.
\newblock \bibinfo{journal}{Sci. Rep.} \bibinfo{volume}{6},
  \bibinfo{pages}{30708}.
\bibitem[{de~Gans and Schubert(2004)}]{de2004inkjet}
\bibinfo{author}{de~Gans, B.J.}, \bibinfo{author}{Schubert, U.S.},
  \bibinfo{year}{2004}.
\newblock \bibinfo{title}{Inkjet printing of well-defined polymer dots and
  arrays}.
\newblock \bibinfo{journal}{Langmuir} \bibinfo{volume}{20},
  \bibinfo{pages}{7789--7793}.
\bibitem[{Ghosh et~al.(2015)Ghosh, Chakraborty, Bhandari, Chakraborty and
  DasGupta}]{ghosh2015effect}
\bibinfo{author}{Ghosh, U.U.}, \bibinfo{author}{Chakraborty, M.},
  \bibinfo{author}{Bhandari, A.B.}, \bibinfo{author}{Chakraborty, S.},
  \bibinfo{author}{DasGupta, S.}, \bibinfo{year}{2015}.
\newblock \bibinfo{title}{Effect of surface wettability on crack dynamics and
  morphology of colloidal films}.
\newblock \bibinfo{journal}{Langmuir} \bibinfo{volume}{31},
  \bibinfo{pages}{6001--6010}.
\bibitem[{Giorgiutti-Dauphin{\'e} and Pauchard(2014)}]{giorgiutti2014elapsed}
\bibinfo{author}{Giorgiutti-Dauphin{\'e}, F.}, \bibinfo{author}{Pauchard, L.},
  \bibinfo{year}{2014}.
\newblock \bibinfo{title}{Elapsed time for crack formation during drying}.
\newblock \bibinfo{journal}{Eur. Phys. J. E.} \bibinfo{volume}{37},
  \bibinfo{pages}{1--7}.
\bibitem[{Giorgiutti-Dauphin{\'e} and Pauchard(2016)}]{giorgiutti2016painting}
\bibinfo{author}{Giorgiutti-Dauphin{\'e}, F.}, \bibinfo{author}{Pauchard, L.},
  \bibinfo{year}{2016}.
\newblock \bibinfo{title}{Painting cracks: A way to investigate the pictorial
  matter}.
\newblock \bibinfo{journal}{J. Appl. Phys.} \bibinfo{volume}{120},
  \bibinfo{pages}{065107}.
\bibitem[{Goehring(2013)}]{goehring2013evolving}
\bibinfo{author}{Goehring, L.}, \bibinfo{year}{2013}.
\newblock \bibinfo{title}{Evolving fracture patterns: columnar joints, mud
  cracks and polygonal terrain}.
\newblock \bibinfo{journal}{Phil. Trans. R. Soc. A.} \bibinfo{volume}{371},
  \bibinfo{pages}{20120353}.
\bibitem[{Groisman and Kaplan(1994)}]{groisman1994experimental}
\bibinfo{author}{Groisman, A.}, \bibinfo{author}{Kaplan, E.},
  \bibinfo{year}{1994}.
\newblock \bibinfo{title}{An experimental study of cracking induced by
  desiccation}.
\newblock \bibinfo{journal}{Europhys. Lett.} \bibinfo{volume}{25},
  \bibinfo{pages}{415--420}.
\bibitem[{Gurrala et~al.(2019)Gurrala, Katre, Balusamy, Banerjee and
  Sahu}]{gurrala2019evaporation}
\bibinfo{author}{Gurrala, P.}, \bibinfo{author}{Katre, P.},
  \bibinfo{author}{Balusamy, S.}, \bibinfo{author}{Banerjee, S.},
  \bibinfo{author}{Sahu, K.C.}, \bibinfo{year}{2019}.
\newblock \bibinfo{title}{Evaporation of ethanol-water sessile droplet of
  different compositions at an elevated substrate temperature}.
\newblock \bibinfo{journal}{Int. J. Heat Mass Transf.} \bibinfo{volume}{145},
  \bibinfo{pages}{118770}.
\bibitem[{Hari~Govindha et~al.(2022)Hari~Govindha, Katre, Balusamy, Banerjee
  and Sahu}]{hari2022counter}
\bibinfo{author}{Hari~Govindha, A.}, \bibinfo{author}{Katre, P.},
  \bibinfo{author}{Balusamy, S.}, \bibinfo{author}{Banerjee, S.},
  \bibinfo{author}{Sahu, K.C.}, \bibinfo{year}{2022}.
\newblock \bibinfo{title}{Counter-intuitive evaporation in nanofluids droplets
  due to stick-slip nature}.
\newblock \bibinfo{journal}{Langmuir} \bibinfo{volume}{38},
  \bibinfo{pages}{15361--15371}.
\bibitem[{He and Darhuber(2019)}]{he2019evaporation}
\bibinfo{author}{He, B.}, \bibinfo{author}{Darhuber, A.A.},
  \bibinfo{year}{2019}.
\newblock \bibinfo{title}{Evaporation of water droplets on photoresist
  surfaces--an experimental study of contact line pinning and evaporation
  residues}.
\newblock \bibinfo{journal}{Colloids Surf. A.} \bibinfo{volume}{583},
  \bibinfo{pages}{123912}.
\bibitem[{Hodges et~al.(2010)Hodges, Ding and Biggs}]{hodges2010influence}
\bibinfo{author}{Hodges, C.S.}, \bibinfo{author}{Ding, Y.},
  \bibinfo{author}{Biggs, S.}, \bibinfo{year}{2010}.
\newblock \bibinfo{title}{The influence of nanoparticle shape on the drying of
  colloidal suspensions}.
\newblock \bibinfo{journal}{J. Colloid Interface Sci.} \bibinfo{volume}{352},
  \bibinfo{pages}{99--106}.
\bibitem[{Karapetsas et~al.(2016)Karapetsas, Sahu and
  Matar}]{karapetsas2016evaporation}
\bibinfo{author}{Karapetsas, G.}, \bibinfo{author}{Sahu, K.C.},
  \bibinfo{author}{Matar, O.K.}, \bibinfo{year}{2016}.
\newblock \bibinfo{title}{Evaporation of sessile droplets laden with particles
  and insoluble surfactants}.
\newblock \bibinfo{journal}{Langmuir} \bibinfo{volume}{32},
  \bibinfo{pages}{6871--6881}.
\bibitem[{Katre et~al.(2021)Katre, Balusamy, Banerjee, Chandrala and
  Sahu}]{katre2021evaporation}
\bibinfo{author}{Katre, P.}, \bibinfo{author}{Balusamy, S.},
  \bibinfo{author}{Banerjee, S.}, \bibinfo{author}{Chandrala, L.D.},
  \bibinfo{author}{Sahu, K.C.}, \bibinfo{year}{2021}.
\newblock \bibinfo{title}{Evaporation dynamics of a sessile droplet of binary
  mixture laden with nanoparticles}.
\newblock \bibinfo{journal}{Langmuir} \bibinfo{volume}{37},
  \bibinfo{pages}{6311--6321}.
\bibitem[{Katre et~al.(2022)Katre, Balusamy, Banerjee and
  Sahu}]{katre2022experimental}
\bibinfo{author}{Katre, P.}, \bibinfo{author}{Balusamy, S.},
  \bibinfo{author}{Banerjee, S.}, \bibinfo{author}{Sahu, K.C.},
  \bibinfo{year}{2022}.
\newblock \bibinfo{title}{An experimental investigation of evaporation of
  ethanol--water droplets laden with alumina nanoparticles on a critically
  inclined heated substrate}.
\newblock \bibinfo{journal}{Langmuir} \bibinfo{volume}{38},
  \bibinfo{pages}{4722--4735}.
\bibitem[{Katre et~al.(2020)Katre, Gurrala, Balusamy, Banerjee and
  Sahu}]{katre2020evaporation}
\bibinfo{author}{Katre, P.}, \bibinfo{author}{Gurrala, P.},
  \bibinfo{author}{Balusamy, S.}, \bibinfo{author}{Banerjee, S.},
  \bibinfo{author}{Sahu, K.C.}, \bibinfo{year}{2020}.
\newblock \bibinfo{title}{Evaporation of sessile ethanol-water droplets on a
  critically inclined heated surface}.
\newblock \bibinfo{journal}{Int. J. Multiphase Flow} \bibinfo{volume}{131},
  \bibinfo{pages}{103368}.
\bibitem[{Kim et~al.(2005)Kim, Ekerdt and Willson}]{kim2005importance}
\bibinfo{author}{Kim, E.K.}, \bibinfo{author}{Ekerdt, J.G.},
  \bibinfo{author}{Willson, C.G.}, \bibinfo{year}{2005}.
\newblock \bibinfo{title}{Importance of evaporation in the design of materials
  for step and flash imprint lithography}.
\newblock \bibinfo{journal}{J. Vac. Sci. Technol. B} \bibinfo{volume}{23},
  \bibinfo{pages}{1515--1520}.
\bibitem[{Kim et~al.(2016)Kim, Boulogne, Um, Jacobi, Button and
  Stone}]{kim2016controlled}
\bibinfo{author}{Kim, H.}, \bibinfo{author}{Boulogne, F.}, \bibinfo{author}{Um,
  E.}, \bibinfo{author}{Jacobi, I.}, \bibinfo{author}{Button, E.},
  \bibinfo{author}{Stone, H.A.}, \bibinfo{year}{2016}.
\newblock \bibinfo{title}{Controlled uniform coating from the interplay of
  marangoni flows and surface-adsorbed macromolecules}.
\newblock \bibinfo{journal}{Phys. Rev. Lett.} \bibinfo{volume}{116},
  \bibinfo{pages}{124501}.
\bibitem[{Kumar et~al.(2023)Kumar, Basavaraj and Satapathy}]{kumar2023effect}
\bibinfo{author}{Kumar, S.}, \bibinfo{author}{Basavaraj, M.G.},
  \bibinfo{author}{Satapathy, D.K.}, \bibinfo{year}{2023}.
\newblock \bibinfo{title}{Effect of colloidal surface charge on desiccation
  cracks}.
\newblock \bibinfo{journal}{Langmuir} \bibinfo{volume}{39},
  \bibinfo{pages}{10249--10258}.
\bibitem[{Lama et~al.(2017)Lama, Basavaraj and Satapathy}]{lama2017tailoring}
\bibinfo{author}{Lama, H.}, \bibinfo{author}{Basavaraj, M.G.},
  \bibinfo{author}{Satapathy, D.K.}, \bibinfo{year}{2017}.
\newblock \bibinfo{title}{Tailoring crack morphology in coffee-ring deposits
  via substrate heating}.
\newblock \bibinfo{journal}{Soft Matter} \bibinfo{volume}{13},
  \bibinfo{pages}{5445--5452}.
\bibitem[{Lama et~al.(2016)Lama, Dugyala, Basavaraj and
  Satapathy}]{lama2016magnetic}
\bibinfo{author}{Lama, H.}, \bibinfo{author}{Dugyala, V.R.},
  \bibinfo{author}{Basavaraj, M.G.}, \bibinfo{author}{Satapathy, D.K.},
  \bibinfo{year}{2016}.
\newblock \bibinfo{title}{Magnetic-field-driven crack formation in an
  evaporated anisotropic colloidal assembly}.
\newblock \bibinfo{journal}{Phys. Rev. E} \bibinfo{volume}{94},
  \bibinfo{pages}{012618}.
\bibitem[{Lama et~al.(2021)Lama, Gogoi, Basavaraj, Pauchard and
  Satapathy}]{lama2021synergy}
\bibinfo{author}{Lama, H.}, \bibinfo{author}{Gogoi, T.},
  \bibinfo{author}{Basavaraj, M.G.}, \bibinfo{author}{Pauchard, L.},
  \bibinfo{author}{Satapathy, D.K.}, \bibinfo{year}{2021}.
\newblock \bibinfo{title}{Synergy between the crack pattern and substrate
  elasticity in colloidal deposits}.
\newblock \bibinfo{journal}{Phys. Rev. E} \bibinfo{volume}{103},
  \bibinfo{pages}{032602}.
\bibitem[{Lama et~al.(2018)Lama, Mondal, Basavaraj and
  Satapathy}]{lama2018cracks}
\bibinfo{author}{Lama, H.}, \bibinfo{author}{Mondal, R.},
  \bibinfo{author}{Basavaraj, M.G.}, \bibinfo{author}{Satapathy, D.K.},
  \bibinfo{year}{2018}.
\newblock \bibinfo{title}{Cracks in dried deposits of hematite ellipsoids:
  Interplay between magnetic and hydrodynamic torques}.
\newblock \bibinfo{journal}{J. Colloid Interface Sci.} \bibinfo{volume}{510},
  \bibinfo{pages}{172--180}.
\bibitem[{Lama et~al.(2023)Lama, Pauchard, Giorgiutti-Dauphine and
  Khawas}]{lama2023salinity}
\bibinfo{author}{Lama, H.}, \bibinfo{author}{Pauchard, L.},
  \bibinfo{author}{Giorgiutti-Dauphine, F.}, \bibinfo{author}{Khawas, S.},
  \bibinfo{year}{2023}.
\newblock \bibinfo{title}{Salinity induced stiffening of drying particulate
  film and dynamic warping}.
\newblock \bibinfo{journal}{Phys. Rev. Mater.} \bibinfo{volume}{7},
  \bibinfo{pages}{025604}.
\bibitem[{Li et~al.(2015)Li, Lv, Li, Qu{\'e}r{\'e} and Zheng}]{li2015coffee}
\bibinfo{author}{Li, Y.}, \bibinfo{author}{Lv, C.}, \bibinfo{author}{Li, Z.},
  \bibinfo{author}{Qu{\'e}r{\'e}, D.}, \bibinfo{author}{Zheng, Q.},
  \bibinfo{year}{2015}.
\newblock \bibinfo{title}{From coffee rings to coffee eyes}.
\newblock \bibinfo{journal}{Soft Matter} \bibinfo{volume}{11},
  \bibinfo{pages}{4669--4673}.
\bibitem[{Lilin and Bischofberger(2022)}]{lilin2022criteria}
\bibinfo{author}{Lilin, P.}, \bibinfo{author}{Bischofberger, I.},
  \bibinfo{year}{2022}.
\newblock \bibinfo{title}{Criteria for crack formation and air invasion in
  drying colloidal suspensions}.
\newblock \bibinfo{journal}{Langmuir} \bibinfo{volume}{38},
  \bibinfo{pages}{7442--7447}.
\bibitem[{Lim et~al.(2009)Lim, Han, Chung, Chung, Ko and
  Grigoropoulos}]{lim2009experimental}
\bibinfo{author}{Lim, T.}, \bibinfo{author}{Han, S.}, \bibinfo{author}{Chung,
  J.}, \bibinfo{author}{Chung, J.T.}, \bibinfo{author}{Ko, S.},
  \bibinfo{author}{Grigoropoulos, C.P.}, \bibinfo{year}{2009}.
\newblock \bibinfo{title}{Experimental study on spreading and evaporation of
  inkjet printed pico-liter droplet on a heated substrate}.
\newblock \bibinfo{journal}{Int. J. Heat Mass Transf.} \bibinfo{volume}{52},
  \bibinfo{pages}{431--441}.
\bibitem[{Lim et~al.(2012)Lim, Yang, Lee, Chung and Hong}]{lim2012deposit}
\bibinfo{author}{Lim, T.}, \bibinfo{author}{Yang, J.}, \bibinfo{author}{Lee,
  S.}, \bibinfo{author}{Chung, J.}, \bibinfo{author}{Hong, D.},
  \bibinfo{year}{2012}.
\newblock \bibinfo{title}{Deposit pattern of inkjet printed pico-liter
  droplet}.
\newblock \bibinfo{journal}{Int. J. Precis. Eng. Manuf.} \bibinfo{volume}{13},
  \bibinfo{pages}{827--833}.
\bibitem[{Liou et~al.(2024)Liou, Hsu, Lin and Wang}]{liou2024suppression}
\bibinfo{author}{Liou, G.F.}, \bibinfo{author}{Hsu, C.C.},
  \bibinfo{author}{Lin, P.W.}, \bibinfo{author}{Wang, P.Y.},
  \bibinfo{year}{2024}.
\newblock \bibinfo{title}{Suppression of the coffee-ring effect by controlling
  the solid particle density}.
\newblock \bibinfo{journal}{Phys. Fluids} \bibinfo{volume}{36},
  \bibinfo{pages}{112114}.
\bibitem[{Liu et~al.(2024)Liu, Liu, Sun, Yu and Ni}]{liu2024formation}
\bibinfo{author}{Liu, X.}, \bibinfo{author}{Liu, M.}, \bibinfo{author}{Sun,
  Y.}, \bibinfo{author}{Yu, S.}, \bibinfo{author}{Ni, Y.},
  \bibinfo{year}{2024}.
\newblock \bibinfo{title}{Formation mechanism of radial and circular cracks
  promoted by delamination in drying silica colloidal deposits}.
\newblock \bibinfo{journal}{Phys. Rev. E} \bibinfo{volume}{110},
  \bibinfo{pages}{034801}.
\bibitem[{Lohani et~al.(2020)Lohani, Basavaraj, Satapathy and
  Sarkar}]{lohani2020coupled}
\bibinfo{author}{Lohani, D.}, \bibinfo{author}{Basavaraj, M.G.},
  \bibinfo{author}{Satapathy, D.K.}, \bibinfo{author}{Sarkar, S.},
  \bibinfo{year}{2020}.
\newblock \bibinfo{title}{Coupled effect of concentration, particle size and
  substrate morphology on the formation of coffee rings}.
\newblock \bibinfo{journal}{Colloids Surf. A.} \bibinfo{volume}{589},
  \bibinfo{pages}{124387}.
\bibitem[{Ma and Wang(2012)}]{ma2012effect}
\bibinfo{author}{Ma, W.}, \bibinfo{author}{Wang, Y.}, \bibinfo{year}{2012}.
\newblock \bibinfo{title}{Effect of salt concentration on the pattern formation
  of colloidal suspension}.
\newblock \bibinfo{journal}{Phys. Procedia} \bibinfo{volume}{24},
  \bibinfo{pages}{122--126}.
\bibitem[{Mailer and Clegg(2014)}]{mailer2014cracking}
\bibinfo{author}{Mailer, A.G.}, \bibinfo{author}{Clegg, P.S.},
  \bibinfo{year}{2014}.
\newblock \bibinfo{title}{Cracking in films of titanium dioxide nanoparticles
  with varying interaction strength}.
\newblock \bibinfo{journal}{J. Colloid Interface Sci.} \bibinfo{volume}{417},
  \bibinfo{pages}{317--324}.
\bibitem[{Mondal and Basavaraj(2019)}]{mondal2019influence}
\bibinfo{author}{Mondal, R.}, \bibinfo{author}{Basavaraj, M.G.},
  \bibinfo{year}{2019}.
\newblock \bibinfo{title}{Influence of the drying configuration on the
  patterning of ellipsoids--concentric rings and concentric cracks}.
\newblock \bibinfo{journal}{Phys. Chem. Chem. Phys} \bibinfo{volume}{21},
  \bibinfo{pages}{20045--20054}.
\bibitem[{Mondal and Basavaraj(2020)}]{mondal2020patterning}
\bibinfo{author}{Mondal, R.}, \bibinfo{author}{Basavaraj, M.G.},
  \bibinfo{year}{2020}.
\newblock \bibinfo{title}{Patterning of colloids into spirals via confined
  drying}.
\newblock \bibinfo{journal}{Soft Matter} \bibinfo{volume}{16},
  \bibinfo{pages}{3753--3761}.
\bibitem[{Mondal et~al.(2023)Mondal, Lama and Sahu}]{mondal2023physics}
\bibinfo{author}{Mondal, R.}, \bibinfo{author}{Lama, H.},
  \bibinfo{author}{Sahu, K.C.}, \bibinfo{year}{2023}.
\newblock \bibinfo{title}{Physics of drying complex fluid drop: Flow field,
  pattern formation, and desiccation cracks}.
\newblock \bibinfo{journal}{Phys. Fluids} \bibinfo{volume}{35},
  \bibinfo{pages}{061301}.
\bibitem[{Mondal et~al.(2018)Mondal, Semwal, Kumar, Thampi and
  Basavaraj}]{mondal2018patterns}
\bibinfo{author}{Mondal, R.}, \bibinfo{author}{Semwal, S.},
  \bibinfo{author}{Kumar, P.L.}, \bibinfo{author}{Thampi, S.P.},
  \bibinfo{author}{Basavaraj, M.G.}, \bibinfo{year}{2018}.
\newblock \bibinfo{title}{Patterns in drying drops dictated by curvature-driven
  particle transport}.
\newblock \bibinfo{journal}{Langmuir} \bibinfo{volume}{34},
  \bibinfo{pages}{11473--11483}.
\bibitem[{Nam et~al.(2012)Nam, Park and Ko}]{nam2012patterning}
\bibinfo{author}{Nam, K.H.}, \bibinfo{author}{Park, I.H.}, \bibinfo{author}{Ko,
  S.H.}, \bibinfo{year}{2012}.
\newblock \bibinfo{title}{Patterning by controlled cracking}.
\newblock \bibinfo{journal}{Nature} \bibinfo{volume}{485},
  \bibinfo{pages}{221--224}.
\bibitem[{Osman et~al.(2020)Osman, Goehring, Stitt and
  Shokri}]{osman2020controlling}
\bibinfo{author}{Osman, A.}, \bibinfo{author}{Goehring, L.},
  \bibinfo{author}{Stitt, H.}, \bibinfo{author}{Shokri, N.},
  \bibinfo{year}{2020}.
\newblock \bibinfo{title}{Controlling the drying-induced peeling of colloidal
  films}.
\newblock \bibinfo{journal}{Soft Matter} \bibinfo{volume}{16},
  \bibinfo{pages}{8345--8351}.
\bibitem[{Pauchard(2006)}]{pauchard2006patterns}
\bibinfo{author}{Pauchard, L.}, \bibinfo{year}{2006}.
\newblock \bibinfo{title}{Patterns caused by buckle-driven delamination in
  desiccated colloidal gels}.
\newblock \bibinfo{journal}{Europhysics letters} \bibinfo{volume}{74},
  \bibinfo{pages}{188}.
\bibitem[{Pauchard and Couder(2004)}]{pauchard2004invagination}
\bibinfo{author}{Pauchard, L.}, \bibinfo{author}{Couder, Y.},
  \bibinfo{year}{2004}.
\newblock \bibinfo{title}{Invagination during the collapse of an inhomogeneous
  spheroidal shell}.
\newblock \bibinfo{journal}{Europhys. Lett.} \bibinfo{volume}{66},
  \bibinfo{pages}{667--673}.
\bibitem[{Prasad et~al.(2014)Prasad, Lin, Rao and
  Seshia}]{prasad2014monitoring}
\bibinfo{author}{Prasad, A.}, \bibinfo{author}{Lin, A.T.H.},
  \bibinfo{author}{Rao, V.R.}, \bibinfo{author}{Seshia, A.A.},
  \bibinfo{year}{2014}.
\newblock \bibinfo{title}{Monitoring sessile droplet evaporation on a
  micromechanical device}.
\newblock \bibinfo{journal}{Analyst} \bibinfo{volume}{139},
  \bibinfo{pages}{5538--5546}.
\bibitem[{Rey et~al.(2022)Rey, Walter, Harrer, Perez, Chiera, Nair, Ickler,
  Fuchs, Michaud, Uttinger et~al.}]{rey2022versatile}
\bibinfo{author}{Rey, M.}, \bibinfo{author}{Walter, J.},
  \bibinfo{author}{Harrer, J.}, \bibinfo{author}{Perez, C.M.},
  \bibinfo{author}{Chiera, S.}, \bibinfo{author}{Nair, S.},
  \bibinfo{author}{Ickler, M.}, \bibinfo{author}{Fuchs, A.},
  \bibinfo{author}{Michaud, M.}, \bibinfo{author}{Uttinger, M.J.}, et~al.,
  \bibinfo{year}{2022}.
\newblock \bibinfo{title}{Versatile strategy for homogeneous drying patterns of
  dispersed particles}.
\newblock \bibinfo{journal}{Nat. Commun.} \bibinfo{volume}{13},
  \bibinfo{pages}{2840}.
\bibitem[{Sanyal et~al.(2015)Sanyal, Basu and
  Chaudhuri}]{sanyal2015agglomeration}
\bibinfo{author}{Sanyal, A.}, \bibinfo{author}{Basu, S.},
  \bibinfo{author}{Chaudhuri, S.}, \bibinfo{year}{2015}.
\newblock \bibinfo{title}{Agglomeration front dynamics: Drying in sessile
  nano-particle laden droplets}.
\newblock \bibinfo{journal}{Chem. Eng. Sci.} \bibinfo{volume}{123},
  \bibinfo{pages}{164--169}.
\bibitem[{Shao et~al.(2020)Shao, Duan, Hou and Zhong}]{shao2020role}
\bibinfo{author}{Shao, X.}, \bibinfo{author}{Duan, F.}, \bibinfo{author}{Hou,
  Y.}, \bibinfo{author}{Zhong, X.}, \bibinfo{year}{2020}.
\newblock \bibinfo{title}{Role of surfactant in controlling the deposition
  pattern of a particle-laden droplet: Fundamentals and strategies}.
\newblock \bibinfo{journal}{Adv. Colloid Interface Sci.} \bibinfo{volume}{275},
  \bibinfo{pages}{102049}.
\bibitem[{Shimobayashi et~al.(2018)Shimobayashi, Tsudome and
  Kurimura}]{shimobayashi2018suppression}
\bibinfo{author}{Shimobayashi, S.F.}, \bibinfo{author}{Tsudome, M.},
  \bibinfo{author}{Kurimura, T.}, \bibinfo{year}{2018}.
\newblock \bibinfo{title}{Suppression of the coffee-ring effect by
  sugar-assisted depinning of contact line}.
\newblock \bibinfo{journal}{Sci. Rep.} \bibinfo{volume}{8},
  \bibinfo{pages}{17769}.
\bibitem[{Singh et~al.(2009)Singh, Bhosale and Tirumkudulu}]{singh2009cracking}
\bibinfo{author}{Singh, K.B.}, \bibinfo{author}{Bhosale, L.R.},
  \bibinfo{author}{Tirumkudulu, M.S.}, \bibinfo{year}{2009}.
\newblock \bibinfo{title}{Cracking in drying colloidal films of flocculated
  dispersions}.
\newblock \bibinfo{journal}{Langmuir} \bibinfo{volume}{25},
  \bibinfo{pages}{4284--4287}.
\bibitem[{Somasundaran et~al.(1995)Somasundaran, Shrotri and
  Ananthapadmanabhan}]{somasundaran1995deposition}
\bibinfo{author}{Somasundaran, P.}, \bibinfo{author}{Shrotri, S.},
  \bibinfo{author}{Ananthapadmanabhan, K.}, \bibinfo{year}{1995}.
\newblock \bibinfo{title}{Deposition of latex particles: Theoretical and
  experimental aspect}.
\newblock \bibinfo{journal}{Miner. Process.: Recent Adv. Future Trends} ,
  \bibinfo{pages}{126--137}.
\bibitem[{Still et~al.(2012)Still, Yunker and Yodh}]{still2012surfactant}
\bibinfo{author}{Still, T.}, \bibinfo{author}{Yunker, P.J.},
  \bibinfo{author}{Yodh, A.G.}, \bibinfo{year}{2012}.
\newblock \bibinfo{title}{Surfactant-induced marangoni eddies alter the
  coffee-rings of evaporating colloidal drops}.
\newblock \bibinfo{journal}{Langmuir} \bibinfo{volume}{28},
  \bibinfo{pages}{4984--4988}.
\bibitem[{Thampi and Basavaraj(2023)}]{thampi2023drying}
\bibinfo{author}{Thampi, S.P.}, \bibinfo{author}{Basavaraj, M.G.},
  \bibinfo{year}{2023}.
\newblock \bibinfo{title}{Drying drops of colloidal dispersions}.
\newblock \bibinfo{journal}{Annu. Rev. Chem. Biomol. Eng.}
  \bibinfo{volume}{14}, \bibinfo{pages}{53--83}.
\bibitem[{Tirumkudulu and Russel(2005)}]{tirumkudulu2005cracking}
\bibinfo{author}{Tirumkudulu, M.S.}, \bibinfo{author}{Russel, W.B.},
  \bibinfo{year}{2005}.
\newblock \bibinfo{title}{Cracking in drying latex films}.
\newblock \bibinfo{journal}{Langmuir} \bibinfo{volume}{21},
  \bibinfo{pages}{4938--4948}.
\bibitem[{Weldon et~al.(2017)Weldon, Joshi, Routh and
  Gilchrist}]{weldon2017uniformly}
\bibinfo{author}{Weldon, A.L.}, \bibinfo{author}{Joshi, K.},
  \bibinfo{author}{Routh, A.F.}, \bibinfo{author}{Gilchrist, J.F.},
  \bibinfo{year}{2017}.
\newblock \bibinfo{title}{Uniformly spaced nanoscale cracks in nanoparticle
  films deposited by convective assembly}.
\newblock \bibinfo{journal}{J. Colloid Interface Sci.} \bibinfo{volume}{487},
  \bibinfo{pages}{80--87}.
\bibitem[{Weon and Je(2010)}]{weon2010capillary}
\bibinfo{author}{Weon, B.M.}, \bibinfo{author}{Je, J.H.}, \bibinfo{year}{2010}.
\newblock \bibinfo{title}{Capillary force repels coffee-ring effect}.
\newblock \bibinfo{journal}{Phys. Rev. E.} \bibinfo{volume}{82},
  \bibinfo{pages}{015305}.
\bibitem[{Xu et~al.(2017)Xu, Hong, Sun, Wang and Tong}]{xu2017effect}
\bibinfo{author}{Xu, G.}, \bibinfo{author}{Hong, W.}, \bibinfo{author}{Sun,
  W.}, \bibinfo{author}{Wang, T.}, \bibinfo{author}{Tong, Z.},
  \bibinfo{year}{2017}.
\newblock \bibinfo{title}{Effect of salt concentration on the motion of
  particles near the substrate in drying sessile colloidal droplets}.
\newblock \bibinfo{journal}{Langmuir} \bibinfo{volume}{33},
  \bibinfo{pages}{685--695}.
\bibitem[{Yan et~al.(2008)Yan, Gao, Sharma, Chiang and Wong}]{yan2008particle}
\bibinfo{author}{Yan, Q.}, \bibinfo{author}{Gao, L.}, \bibinfo{author}{Sharma,
  V.}, \bibinfo{author}{Chiang, Y.M.}, \bibinfo{author}{Wong, C.},
  \bibinfo{year}{2008}.
\newblock \bibinfo{title}{Particle and substrate charge effects on colloidal
  self-assembly in a sessile drop}.
\newblock \bibinfo{journal}{Langmuir} \bibinfo{volume}{24},
  \bibinfo{pages}{11518--11522}.
\bibitem[{Yanagisawa et~al.(2014)Yanagisawa, Sakai, Isobe, Matsushita and
  Nakajima}]{yanagisawa2014investigation}
\bibinfo{author}{Yanagisawa, K.}, \bibinfo{author}{Sakai, M.},
  \bibinfo{author}{Isobe, T.}, \bibinfo{author}{Matsushita, S.},
  \bibinfo{author}{Nakajima, A.}, \bibinfo{year}{2014}.
\newblock \bibinfo{title}{Investigation of droplet jumping on superhydrophobic
  coatings during dew condensation by the observation from two directions}.
\newblock \bibinfo{journal}{Appl. Surf. Sci.} \bibinfo{volume}{315},
  \bibinfo{pages}{212--221}.
\bibitem[{Yu et~al.(2021)Yu, Kadir, Liu and Huang}]{yu2021droplet}
\bibinfo{author}{Yu, Z.}, \bibinfo{author}{Kadir, M.}, \bibinfo{author}{Liu,
  Y.}, \bibinfo{author}{Huang, J.}, \bibinfo{year}{2021}.
\newblock \bibinfo{title}{Droplet-capturing coatings on environmental surfaces
  based on cosmetic ingredients}.
\newblock \bibinfo{journal}{Chem} \bibinfo{volume}{7},
  \bibinfo{pages}{2201--2211}.
\bibitem[{Yunker et~al.(2011)Yunker, Still, Lohr and
  Yodh}]{yunker2011suppression}
\bibinfo{author}{Yunker, P.J.}, \bibinfo{author}{Still, T.},
  \bibinfo{author}{Lohr, M.A.}, \bibinfo{author}{Yodh, A.},
  \bibinfo{year}{2011}.
\newblock \bibinfo{title}{Suppression of the coffee-ring effect by
  shape-dependent capillary interactions}.
\newblock \bibinfo{journal}{Nature} \bibinfo{volume}{476},
  \bibinfo{pages}{308--311}.
\bibitem[{Zhang et~al.(2022)Zhang, Li and Wang}]{zhang2022ultrafast}
\bibinfo{author}{Zhang, C.}, \bibinfo{author}{Li, W.}, \bibinfo{author}{Wang,
  Y.}, \bibinfo{year}{2022}.
\newblock \bibinfo{title}{Ultrafast self-assembly of colloidal photonic
  crystals during low-pressure-assisted evaporation of droplets}.
\newblock \bibinfo{journal}{J. Phys. Chem. Lett.} \bibinfo{volume}{13},
  \bibinfo{pages}{3776--3780}.

\end{thebibliography}

\end{document}